\documentclass[12pt]{article}

\usepackage{amsmath}
\usepackage{amssymb}
\usepackage{graphicx}
\usepackage[letterpaper,margin=1in]{geometry}
\usepackage{longtable}
\usepackage{booktabs}
\usepackage{ragged2e}
\usepackage{array}
\usepackage{xurl}
\usepackage{adjustbox}
\usepackage{threeparttable}
\usepackage{hyperref}
\usepackage{url}
\usepackage{soul}
\usepackage{caption}
\date{}

\title{\textbf{Scientific Discovery in the Age of \\ AI and Supercomputing \\}}

\author{
	Stefano Bianchini$^{1}\thanks{Corresponding author; e-mail: s.bianchini@unistra.fr}$ ,
	Aldo Geuna$^{2,3}$,
	Fazliddin Shermatov$^{1,4}$\\[0.3em]
	\small$^{1}$BETA, Université de Strasbourg, France\\
	\small$^{2}$Department of Cultures, Politics and Society, University of Turin, Italy\\
	\small$^{3}$Collegio Carlo Alberto, Turin, Italy\\
	\small$^{4}$Université Sorbonne Paris Nord, France
}

\begin{document} 

\maketitle

\vspace{-1em}

\begin{abstract}

\noindent Artificial intelligence (AI) and high-performance computing (HPC) are transforming scientific capabilities and the way science is conducted. Yet their combined impact on scientific discovery remains poorly understood, as do inequalities in access to these capabilities across countries and institutions. Drawing on metadata from more than five million scientific publications (2000–2024) across 27 fields, we examine how the convergence of AI and HPC correlates with scientific breakthroughs. Our results show that this computational synergy is most pronounced at the scientific frontier: research combining AI and HPC is more likely to introduce novel ideas and achieve top-cited status than either conventional work or research using AI or HPC in isolation. We also document growing disparities in access to supercomputing resources and AI expertise, which are increasingly concentrated in a small number of regions (dominated by the United States and China, though the EU27 aggregate maintains high competitiveness in combined AI+HPC output). The future of discovery will therefore depend not only on advances in algorithms and computing power, but also on enacting policies that democratise these capabilities across the global scientific ecosystem.
\end{abstract}

\newpage

\section*{Introduction}

\noindent In recent decades, a few signs of fatigue have begun to emerge in the scientific enterprise. Despite the rapid growth in publications and data, progress seems to be slowing down. Some evidence suggests that ideas are becoming harder to find \cite{Bloom2020}, research outputs are becoming less disruptive \cite{Park2023}, and productivity in research and innovation has declined across most fields \cite{Brynjolfsson2017}. This apparent deceleration has revived an age-old concern: that science may be about to enter a phase of diminishing returns, the most accessible problems having already been solved \cite{Cowen2011}, while the complexity of knowledge continues to rise \cite{Jones2009}.

Yet just as science may be facing diminishing returns, a new class of computational technologies has begun to transform research practices. In recent years, artificial intelligence (AI) and high-performance computing (HPC) have underpinned many of the most remarkable advances made. AI models trained on vast scientific datasets are capable of identifying regularities and generating hypotheses that would have otherwise escaped human reasoning, while HPC systems enable model training, testing, and large-scale simulations to be carried out. From the prediction of protein structures \cite{Jumper2021} to the discovery of new materials \cite{Merchant2023} and the control of fusion plasma \cite{Degrave2022}, the combination of AI and supercomputing is redefining what science can achieve.

This convergence has not happened overnight. Earlier waves of information and communication technologies (ICTs) had already accelerated data collection and analysis, but they did so without fundamentally altering the logic of scientific inference \cite{Mokyr2015}. AI and HPC, by contrast, have shifted science from automation to autonomy: from machines that assist computation to systems that learn, generalize, and explore enormous combinatorial spaces \cite{Krenn2022}. In the process, they have blurred the boundaries between theory, modeling, and experimentation, creating new forms of “computational empiricism” in which discovery emerges as much from pattern recognition as from deductive reasoning.

AI continues to redefine performance boundaries and, in its own quiet way, to reshape the way science gets done. No longer simply a tool for data crunching, it now permeates virtually every step in the scientific pipeline, from hypothesis generation to experimental design to publication \cite{Gao2024}. For some, deep learning represents much more than mere progress in computation; it is a new engine of invention, redefining how science operates and where its current frontier stands \cite{Cockburn2018, Bianchini2022} -- even if contemporary GenAI systems may still struggle to generate novel discoveries from scratch \cite{Ding2025}. In 2024, the Nobel Prizes in Chemistry and Physics recognized breakthroughs made possible by AI, with DeepMind capturing the prevailing sentiment with its forecast that we may be entering “\textit{a new golden age of discovery}” \cite{Griffin2024}.

Behind the emergence of AI lies an equally important, if less celebrated, revolution in hardware. The massive gains in model performance have not resulted from better algorithms alone, but rather from the development of ever more powerful computers. Moore’s Law may be stalling, yet computing power continues to grow, fuelled by the global arms race in silicon. Nations are investing hundreds of billions of dollars in supercomputing, and semiconductors have become one of the world’s most coveted trade commodities \cite{Miller2022}. The European Union alone has committed roughly €7 billion (\$7.5 billion) to its EuroHPC Joint Undertaking (2021–2027) to build a world-class supercomputing ecosystem \cite{EuroHPC}. Meanwhile, the United States and China continue to push back the frontier, in a race to develop and deploy successive generations of exascale-class systems (albeit that fewer details have been revealed about their consolidated budgets) – an escalating competition that has come to define the infrastructure of modern science.

These trends signal the emergence of a new mode of discovery and raise a fundamental question: Can the convergence of AI and HPC actually deliver the leap forward that science has been waiting for? In this article, we provide statistical evidence for a strong correlation between the joint use of AI and HPC and frontier scientific discovery -- which we operationalise as research that is simultaneously highly novel and highly impactful. We show that a growing share of high-impact, high-novelty research relies upon these two technologies, a trend that cuts across nearly every scientific discipline. But the shift also raises a number of red flags. Indeed, as access to supercomputing and AI expertise becomes concentrated in a handful of countries \cite{Schmallenbach, AlShebli}, elite institutions, and private firms \cite{Ahmed2023, Jurowetzki2026} (see also \hyperref[fig:3]{Figure 3}), there is a growing risk of a more stratified scientific ecosystem, one in which a few players increasingly determine the direction of discovery. We return to the implications for policy and for the scientific community in the concluding section.

\section*{Results}
\subsection*{The convergence of AI and HPC in scientific research}

We compiled metadata for over five million AI-related publications, drawing on a curated list of 203 keywords to search titles, abstracts, and author keywords in Scopus. The results are robust to alternative, more restrictive AI keyword definitions [see \hyperref[sm:data_construction]{Supplementary Materials (SM)}]. To detect the use of high-performance computing (HPC), we mined acknowledgment sections for explicit mentions of computational resources (terms such as \textit{“HPC”} or \textit{“supercomputer”}) -- a strategy that reliably captures the actual use of large-scale computing in scientific research. This approach was validated through case studies conducted in collaboration with HPC facilities at the Universities of Strasbourg and Turin [see \hyperref[sm:data_validation]{SM}]. In total, we identified over 200,000 publications acknowledging HPC use. Papers that fell at the intersection of our AI and HPC corpora were classified as AI+HPC, denoting research that explicitly combines the two technologies. 

AI+HPC research can take two broad forms: the use of HPC to either train AI models or to run large-scale inferences. To better understand this intersection, we applied a Large Language Model (LLM)-based classifier to full-text papers (when available) [see \hyperref[sm:data_classification]{SM}], distinguishing between two categories. The first, AI-development, refers to research that builds or improves AI models, algorithms, or theoretical foundations. The second, AI-use, includes studies that apply existing AI tools and techniques to problems in other domains. Representative examples of each category are provided in the \hyperref[tab:use_development]{SM}.

\begin{figure}[ht!]
    \centering
    \includegraphics[width=\textwidth]{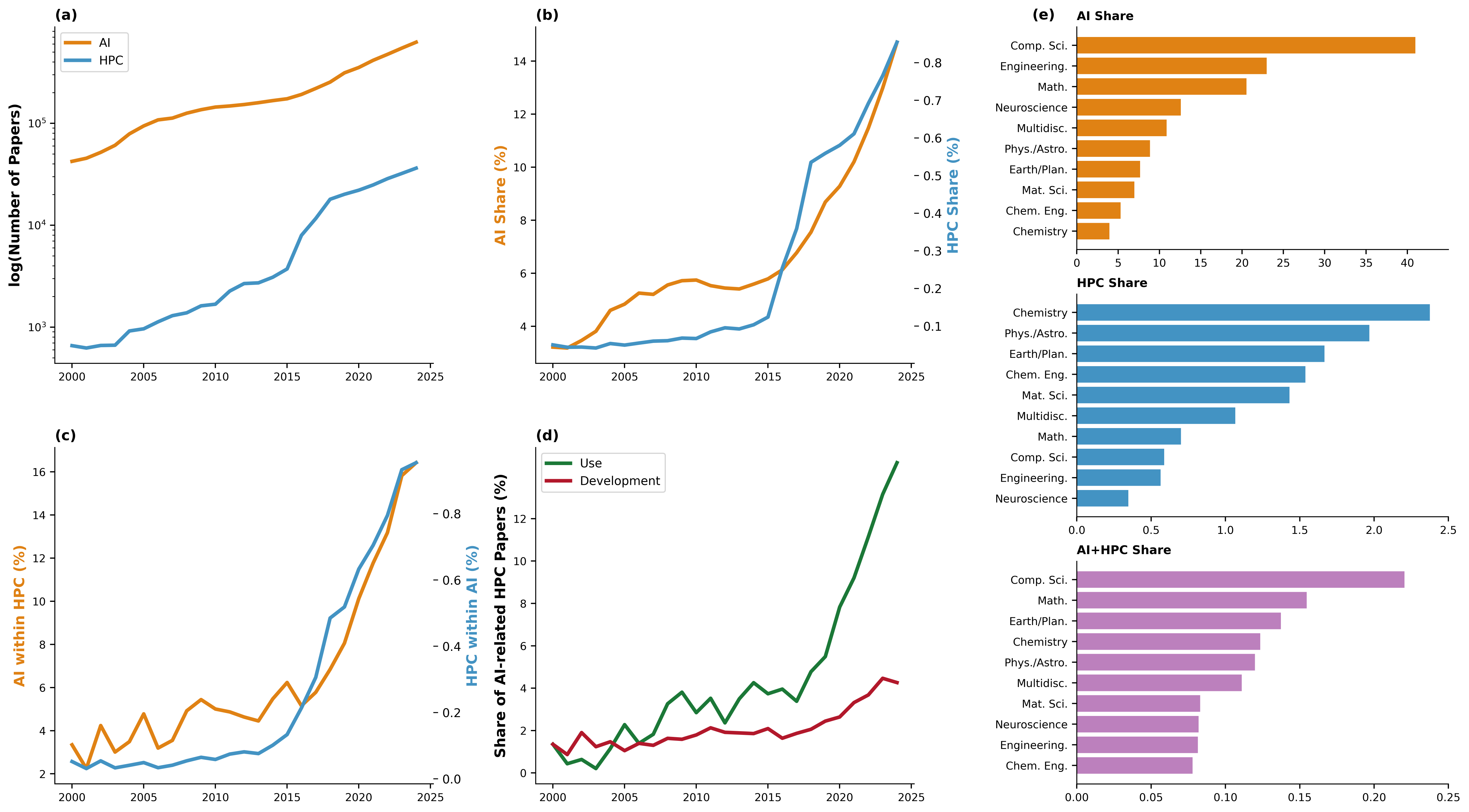} 
    \caption{\textbf{Growth and convergence of AI and HPC in scientific research.}
    (\textbf{A}) Number of AI and HPC publications over time (log scale). 
    (\textbf{B}) Share of all scientific papers involving AI (orange, left axis) and HPC (blue, right axis), based on Scopus data. 
    (\textbf{C}) Co-occurrence of AI and HPC publications: share of HPC papers that use AI (orange, left axis) and share of AI papers that use HPC (blue, right axis). 
    (\textbf{D}) Among AI+HPC papers, classification by contribution type: “AI use” (green) vs. “AI development” (red). 
    (\textbf{E}) Disciplinary distribution of AI, HPC, and combined AI+HPC work. For each field, bars show the share of total output involving AI (top, orange), HPC (middle, blue), or both (bottom, purple).}
    \label{fig:1}
\end{figure}

The growth in AI research over the past two decades has been nothing short of exponential (\hyperref[fig:1]{Figure 1a}). In 2000, there were roughly 40,000 AI publications; by 2024, that number exceeded 600,000. Computer Science accounts for 57\% of all AI publications, and together with Engineering and Mathematics, these fields make up over three-quarters of total AI output. Yet AI’s reach is widening rapidly. Since 2000, its compound annual growth rate has exceeded 8\% across all disciplines. As \textit{Nature} observed a few years ago: “\textit{Few fields are untouched by the machine-learning revolution, from materials science to drug discovery, quantum physics to medicine}” \cite{Nature2019}. The recent surge in generative AI (GenAI) has further accelerated this diffusion. Perhaps the most telling, and most ironic, nod to AI’s growing role in science came when Nature’s 2023 “Ten People Who Shaped Science” list featured, for the first time, a non-human: ChatGPT. As the editorial put it, this “\textit{poster child for generative AI software represents a potential new era for science}” \cite{Nature2023}. 

A similar expansion is evident in research powered by HPC. In 2024, just under 1\% of all scientific papers acknowledged the use of HPC – still only a small fraction, but a six-fold increase on 2015 figures, when growth began to take off (\hyperref[fig:1]{Figure 1b}). HPC use remains concentrated in the core STEM disciplines: Chemistry, Physics and Astronomy, Materials Science, Engineering, and Computer Science together account for just over three-quarters of all HPC-tagged publications (\hyperref[fig:1]{Figure 1e}). Yet adoption is rising sharply elsewhere, even in disciplines traditionally seen as less computationally intensive. 

Most striking, however, is the rapid convergence of AI and HPC in research workflows (\hyperref[fig:1]{Figure 1c}). By 2024, roughly 16\% of HPC-related publications incorporated some AI methods, while about 1\% of AI papers explicitly acknowledged HPC use. This imbalance might suggest that, while HPC is essential for heavy-duty AI, much of AI research still runs on more modest computational infrastructure. Since around 2017, the fastest growth has been recorded in applied work (\hyperref[fig:1]{Figure 1d}): researchers are increasingly deploying AI to tackle domain-specific challenges. As of 2024, almost 15\% of HPC-tagged papers integrate AI for applied purposes, compared to just 4\% focused on advancing AI itself. The synergy is powerful. As one Microsoft blog post put it: “\textit{If you think of discovery as a funnel, we used AI to widen the mouth of the funnel. If we used only HPC, we could look at maybe 1,000 candidates. When we put AI at the top of the funnel to screen out all but the best candidates, we could look at 32-and-a-half million in 80 hours. That combination is really powerful}” \cite{FedScoop2024}.

\subsection*{AI and HPC in frontier scientific discovery}

\begin{figure}[ht!]
    \centering
    \includegraphics[width=\textwidth]{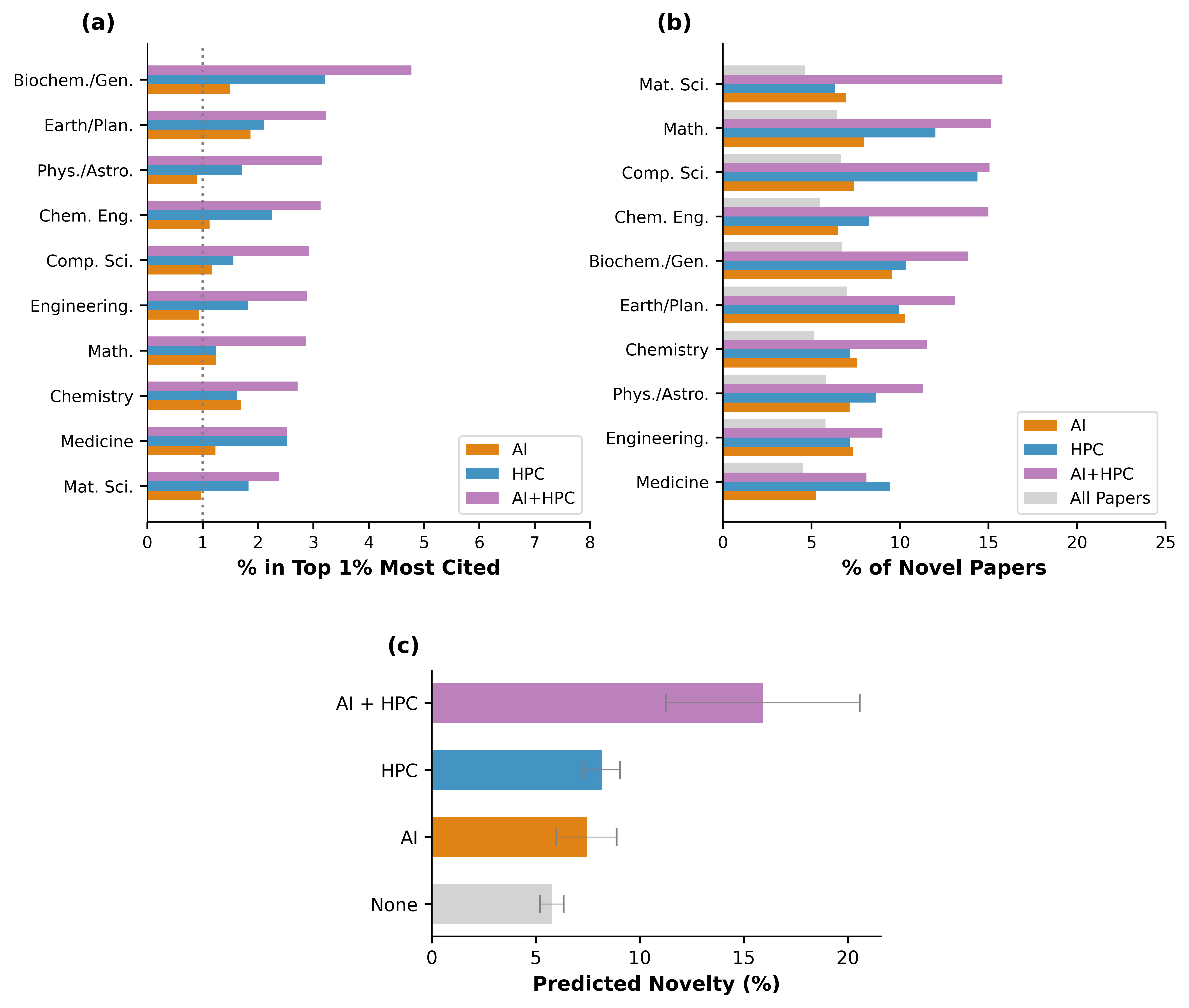}
    \caption{\textbf{Breakthrough potential in AI and HPC-powered science.}
    (\textbf{A}) Share of papers in the top 1\% of citations by field and computational approach: AI only (orange), HPC only (blue), and both AI and HPC (purple). The dashed line marks the 1\% baseline expected by random chance. 
    (\textbf{B}) Share of novel papers (i.e., those introducing a new term subsequently reused) among the top 1\% most cited, across fields and computational approaches. Panels A and B display the ten scientific fields with the largest volume of publications combining AI and HPC.
    (\textbf{C}) Predicted probability of novelty (from regression models controlling for confounders such as affiliation, prior citations, and year) for four groups: papers using neither AI nor HPC (gray), AI only (orange), HPC only (blue), and both (purple).}
    \label{fig:2}
\end{figure}

A long tradition in innovation studies shows that output scales with available resources \cite{Lepori2019}. Access to data, computing infrastructure, and general-purpose technologies such as Information Technology (IT) tends to boost scientific productivity \cite{McKeon2025}. Yet productivity is not synonymous with breakthrough quality. Here, we focus on frontier scientific discovery, defined as research that is both highly impactful and novel. We operationalise these dimensions using two indicators: papers that (i) rank among the top 1\% most cited (\textit{impact}) and (ii) introduce at least one new term (i.e., words or multi-word expressions) subsequently reused in later work (\textit{novelty}). Full methodological details are provided in the \hyperref[sm:methodology]{SM}.

To assess whether AI and HPC are correlated with scientific breakthroughs, we examine how often papers using these tools land among the top 1\% most cited (\hyperref[fig:2]{Figure 2a}). On their own, HPC papers are more likely to reach this threshold than AI-only work – a pattern that holds across nearly all disciplines. However, the real payoff is delivered when the two are used together: AI+HPC papers consistently show the highest odds of being top-cited, far exceeding random expectations. In Biochemistry, Genetics \& Molecular Biology, for example, 5\% of AI+HPC papers enter the top 1\% – five times the baseline. In other words, while any paper has a 1\% chance of making this cut, AI+HPC papers in this domain are five times more likely to do so. Across most other disciplines, this rate is also high, exceeding 2\%.

The same holds for novelty (\hyperref[fig:2]{Figure 2b}). Among top-cited papers, we measured how often they introduce genuinely new terms. AI, HPC, and especially their combination all correlate with a higher propensity to introduce novel terminology. Once again, the greatest boost is recorded when the two are used together. In eight out of 27 fields, more than 10\% of AI+HPC papers qualify as novel – roughly double the rate observed for work using neither technology.

To account for confounding factors (e.g., author affiliation and other researcher-level indicators such as prior citations), we estimated statistical models [see \hyperref[sm:methodology]{SM}]. In the baseline group (papers using neither AI nor HPC), the predicted probability of introducing a novel term is roughly 5\%. Using either AI or HPC alone is associated with a modest increase of 7–8\%. But when the two are used together, the predicted likelihood jumps to 16\% – a three-fold increase beyond additive expectations. These findings are robust to alternative novelty measures, alternative AI keyword definitions, the exclusion of Computer Science and other individual subject areas as well as other sample restrictions [see \hyperref[sm:methodology]{SM}].

\begin{figure}[htbp]
    \centering
    \includegraphics[width=\textwidth]{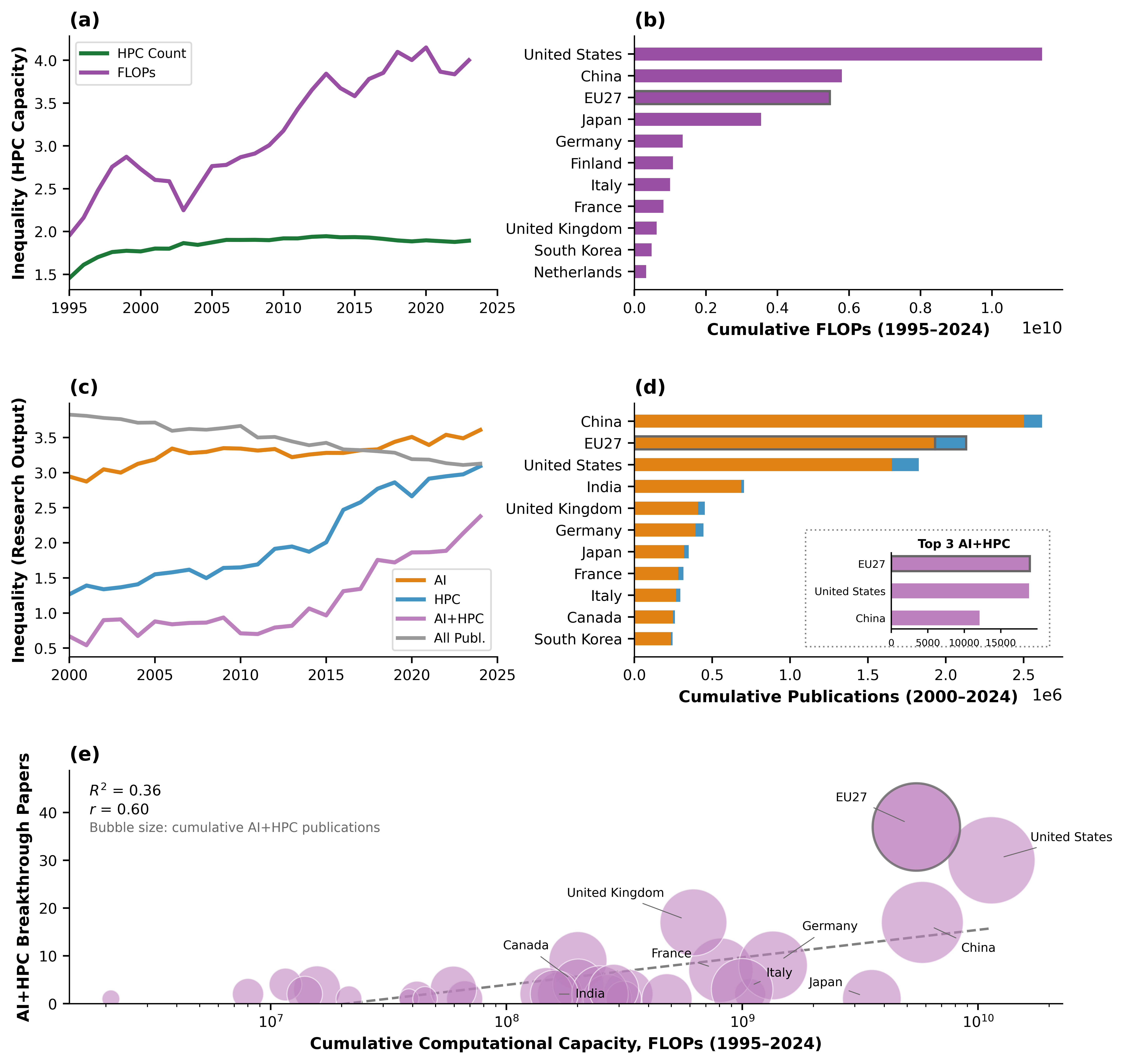}
\caption{\textbf{Global inequality in computational resources and AI/HPC-powered science.} 
(\textbf{A}) Inequality in the global distribution of high-performance computing resources, measured using the Mean Log Deviation (MLD) for HPC system counts and FLOPs (1995–2024), based on the Top500 dataset (own elaboration, see \hyperref[sm:computing_capacity]{SM}).
(\textbf{B}) Top 10 countries and the EU27 by cumulative supercomputing capacity (FLOPs), 1995--2024.
(\textbf{C}) Inequality in scientific publications for AI, HPC, AI+HPC, and all research fields, measured using the Mean Log Deviation (MLD), 2000--2024.
(\textbf{D}) Top 10 countries and the EU27 by cumulative AI and HPC publications, 2000--2024. The inset reports the top 3 countries or regions in cumulative AI+HPC publications. 
(\textbf{E}) Association between cumulative computational capacity (1995–2024) and the number of breakthrough AI+HPC publications (2017--2024). Each bubble represents a country (or the EU27 aggregate) with non-zero cumulative FLOPs, at least 25 AI+HPC publications, and at least one breakthrough AI+HPC publication; bubble size is proportional to cumulative AI+HPC publications. The dashed line reports the fitted semi-log relationship, with the corresponding $R^2$ and Pearson correlation coefficient $r$ shown in the panel.}
\label{fig:3}
\end{figure}

\section*{Discussion}

In this paper, we provide evidence that AI and HPC, particularly when used in combination (AI+HPC), are strongly associated with frontier scientific discovery. This finding is robust across a wide range of sensitivity analyses, but some limitations should be acknowledged. First, our observational analysis cannot establish direct causality; more capable researchers or elite institutions may simply be earlier adopters of these technologies. Second, our keyword-based HPC detection method yields a conservative lower-bound estimate, capturing a 62–70\% recall rate compared to full-text LLM validation. Reassuringly, the validation exercises [see \hyperref[sm:data_validation]{SM}] reveal no systematic differences in field composition, citation impact, or novelty between detected and missed papers. Finally, because the total volume of AI+HPC papers remains relatively small, estimates for this group are necessarily less precise; although the estimated association is consistently large across specifications, it should be interpreted with appropriate caution given the wider confidence intervals (\hyperref[fig:2]{Figure 2c}). With these limitations in mind, we now discuss our findings in the broader context of the rapid growth and increasing concentration of global computing capacity.

As scientific challenges grow more complex, access to cutting-edge computational tools is no longer a luxury but a prerequisite for frontier research. The compute needed to train and operate state-of-the-art AI models is growing rapidly – doubling every five months, according to estimates \cite{EpochAI}. The problem though is not the available capacity, but rather where that capacity is located and who can use it. Science has always been a stratified system \cite{Jones2008} but the imbalance is deepening as industry and elite institutions tighten their grip on the most transformative technologies (AI and HPC included) \cite{Ahmed2023}. Drawing on data from the Top500 list [see \hyperref[sm:computing_capacity]{SM}], we find that supercomputing capacity is clustering within a handful of countries (\hyperref[fig:3]{Figure 3a-b}). Emerging economies barely register. The divide is even starker for AI-specific systems: as of April 2025, the United States accounts for roughly 75\% of global AI supercomputing power, followed by China at around 15\%. Industry dominance is also accelerating, with AI supercomputers in the private sector now outpacing those in academia and government \cite{Pilz2025}. A similar pattern emerges in AI and HPC research itself: the global distribution of publications is highly skewed, and inequality between countries has risen sharply over the past decade, even though overall research output in other scientific domains has remained comparatively stable or slightly less concentrated (\hyperref[fig:3]{Figure 3c–d}). Growth in AI- and HPC-, and AI+HPC-related publications has likewise been concentrated in a small number of scientific systems, led by China and the United States [see \hyperref[sm:figure_sm4]{SM}]. While computational capacity remains heavily concentrated in these two countries, the EU27 aggregate maintains high competitiveness, establishing a strong presence in AI+HPC scientific output (\hyperref[fig:3]{Figure 3d–e}). Moreover, countries with greater computational capacity consistently produce more breakthrough publications (\hyperref[fig:3]{Figure 3e}). Together, these findings suggest that the scientific opportunities created by AI and HPC are increasingly concentrated in countries with the greatest computational resources, raising concerns that unequal access to compute may reinforce existing disparities in scientific leadership.

This concentration raises uncomfortable questions. If these technologies are today so critical in advancing the frontiers of science they will shape which research problems attract funding, which applications mature first, and who ultimately reaps the economic and strategic benefits. Research addressing local or regional challenges may struggle to compete for funding if it does not align with the agendas of better-resourced labs. In the private sector, priorities often lean towards short-term commercial interests rather than long-term public benefits. Left unchecked, we risk drifting towards a two-tier scientific system, one in which a small circle of actors generates frontier knowledge and captures the rewards, while the rest are left to adopt, adapt, or simply watch from the sidelines.

This trajectory is not inevitable. But policy must move quickly and decisively. In Europe, for example, the \textit{Strategy for AI in Science} (launched on October 8, 2025) \cite{EU2025} seeks to accelerate the responsible uptake of AI while reducing disparities in access to AI and HPC through shared computing infrastructure and AI/HPC centres supporting academic researchers and under-resourced institutions. Infrastructure alone, however, will not suffice. Real-world gains will also depend on overcoming bottlenecks in data access, experimentation, and human oversight \cite{Hebenstreit2026}. This requires sustained investment in research capacity through competitive funding, collaboration between leading and emerging institutions, interoperable research infrastructures, and the development of AI and HPC skills across all career stages. For interested readers, a full set of policy recommendations – structured around three pillars: infrastructure, talent, and governance – is available in \cite{EC2025}.

These steps, we believe, must be taken to ensure that the scientific community is prepared for discovery in the age of AI and supercomputing. Only then can science fully harness the potential of these technologies for the benefit of society.

\clearpage
\section*{Methods}

Below we briefly report the methods used in this study. Full methodological details, robustness checks, and validation procedures are available in the \hyperref[sm:data_construction]{Supplementary Materials}.  \\

\noindent \textbf{Data sources}. 
We constructed a large-scale dataset combining bibliometric and textual information on scientific publications indexed in \textit{Scopus} between 2000 and 2024. The sample comprises over 5.3 million records across 27 scientific fields, classified according to the All Science Journal Classification (ASJC) system. For each publication, we extracted metadata on authors, affiliations, journals, citations, abstracts, and (where available) acknowledgments. \\

\noindent \textbf{Identification of AI and HPC publications}.
To identify AI-related research, we used a curated set of 203 keywords referring to machine learning, deep learning, natural language processing, and related sub-fields (e.g., “\textit{neural network}”, “\textit{reinforcement learning}”, “\textit{computer vision}”). We also tested three alternative, more restrictive AI keyword definitions. The first uses a small set of core, unambiguous keywords (e.g., “\textit{artificial intelligence}”, “\textit{machine learning}”, “\textit{deep learning)}”. The second is based on the ten most frequent AI-related keywords in our corpus. The third relies on a methodology-specific dictionary comprising twenty terms [see SM]. Keywords were matched in titles, abstracts, and author keywords using case-insensitive regular expressions. As of March 2025, this procedure yielded 5,193,154 AI publications for the 15-year period 2000 to 2024.  To identify high-performance computing (HPC), we text-mined acknowledgment sections, in which researchers typically credit the computing resources used. Here, we defined our keywords narrowly, including: “\textit{HPC}”, “\textit{high performance comput-er/-ing}”, “\textit{supercomputing}”, and “\textit{supercomputer}”. As of February 2025, this procedure yielded 227,800 HPC papers for the period 2000 to 2024. Publications containing at least one AI keyword were tagged as AI; those with HPC acknowledgments were tagged as HPC. The intersection of these two corpora was labelled AI+HPC and comprised 25,160 papers. \\

\newpage

\noindent \textbf{Validation of HPC detection.} We validated the acknowledgment-based HPC detection method using case studies at two facilities: the University of Strasbourg and the University of Turin. At each facility, we matched registered users to their Scopus author IDs, retrieved all corresponding publications, and analysed the full text of a representative subsample using an LLM-based classifier (Gemini 2.0 Flash) to detect explicit mentions of HPC use. Comparing these classifications with our acknowledgment-based detections yielded recall rates of 62–70\% with no false positives. We found no systematic bias across disciplines or publication years. Acknowledgment-based detection may underestimate HPC use in fields where acknowledgments are uncommon (e.g., the humanities). Similarly, keyword searches cannot fully capture implicit or generic mentions of AI. Our estimates, therefore, likely represent a conservative lower bound on the true prevalence of AI and HPC use in science. \\

\noindent \textbf{Classification of AI–HPC research: AI use vs. development.} 
To characterise how AI and HPC intersect in scientific practice, we analysed the full text of more than 120,000 accessible papers (covering 51\% of the HPC dataset) using LLMs (Gemma3-27B and Gemma2-27B) to detect AI-related content within HPC publications. We distinguished between two categories of AI research: (i) \textit{AI development}, work aimed at creating or improving AI methods, algorithms, models, or theoretical foundations, and (ii) \textit{AI use}, work applying existing AI tools or models to problems in other scientific fields. \\

\noindent \textbf{Measuring novelty and impact.} 
Scientific breakthroughs were assessed along two complementary dimensions: \textit{novelty} and \textit{impact}. Novelty captures linguistic innovation, namely, the introduction of a new term or phrase (1–3-gram) that subsequently diffuses through later publications. The indicator builds on the approach of \cite{Arts2025}, which tokenised titles and abstracts using the \texttt{spaCy} library (v3.7) to identify expressions appearing for the first time in year t in the OpenAlex database. We matched these novelty metrics to our dataset using each publication's DOI. Here, we extend the definition by requiring that a new term appear in at least five subsequent publications within five years, to reduce noise from spurious or idiosyncratic terminology. Highly cited publications were identified as those within the top 1\% of citations in their field and publication year, based on Scopus citation distributions. Citation counts were normalised by field and year to account for disciplinary differences in citation practices. \\

\noindent \textbf{Statistical modelling.} 
We modelled the probability that a paper exhibits high impact or novelty as a function of its computational approach (AI only, HPC only, AI+HPC, or neither), controlling for potential confounders. Probit models were estimated with robust standard errors clustered at the journal level, country, time, and field fixed effects. Control variables included the number of authors, international co-authorship, institutional type (academic, corporate, or governmental), institutional prestige (Shanghai ranking), journal prestige (SCImago Journal Rank) and prior citation stock of the authors. To test for complementarity between AI and HPC, we included the interaction term (\(AI_i \times HPC_i\)) and compared its marginal effect with the sum of individual effects. Our findings are robust to alternative novelty measures (i.e., new phrases, new word and phrase combinations, and semantic distance), alternative AI keyword definitions, the exclusion of individual subject areas (Computer Science, Engineering, Physics), additional sample restrictions (e.g., excluding highly cited papers or those with many authors), and estimation by region [see \hyperref[sm:methodology]{SM}]. \\

\noindent \textbf{Measuring inequality in compute capacity.} 
To assess global disparities in computational resources, we used the \textit{Top500} list of supercomputers (1993–2025) as a proxy for national compute capacity. The performance of each system is measured by the number of floating-point operations per second (FLOPs) that its processors can perform. Aggregating these benchmark results provides total and cumulative capacities by country. Using these data, we constructed a panel covering 56 countries. Although the number of countries hosting at least one Top500 machine has increased, the bulk of compute power remains concentrated within a small elite group. To quantify this inequality, we calculated the \textit{Generalised Entropy Index}, which emphasises deviations around the mean.  \\

\newpage



\newpage

\noindent \textbf{Acknowledgments.} We thank the participants in the Policy Support Facility (PSF) Mutual Learning Exercise (MLE) on National Policies for AI in Science, funded under Horizon Europe and carried out between April 2024 and June 2025, for their insightful discussions and comments. We also thank participants at STI Germany 2024, WOEPS 2025, WINGS 2025, and the International Schumpeter Society Conference 2026, as well as at various seminars, for useful comments. We are especially grateful to Marco Aldinucci and Pierre Gerhard for their valuable feedback during the interviews. We further acknowledge the use of the HPC4AI supercomputing facility. The authors declare that they have no conflict of interest. \\

\noindent \textbf{Funding.} The research leading to the results of this paper has received financial support from the French National Research Agency: SEED (Sustainable Economies in Era of Digital Transformation) – ANR-22-CE26-0013-01 -- and from the Cluster IA Grand Est ENACT. \\

\noindent \textbf{Author contribution.} S.B.: Conceptualization, Methodology, Formal analysis, Visualization, Writing – Original Draft, Supervision, Funding acquisition. A.G.: Conceptualization, Methodology, Investigation, Writing – Review and Editing, Supervision. F.S.: Conceptualization, Methodology, Investigation, Data curation, Formal analysis, Visualization, Writing – Original Draft. \\

\noindent \textbf{Data and code availability.} 
Raw bibliometric data obtained from \textit{Scopus} (Elsevier) are subject to license restrictions and cannot be publicly shared. Aggregate data derived from these records, together with all scripts used for data retrieval, processing, and analysis, are available in our public \href{https://github.com/merlino-sb/AI_HPC}{GitHub repository}. This repository reproduces all analyses reported in the paper and Supplementary Materials.

\clearpage

\newpage

\renewcommand{\thefigure}{S\arabic{figure}}
\renewcommand{\thetable}{S\arabic{table}}
\renewcommand{\theequation}{S\arabic{equation}}
\renewcommand{\thepage}{S\arabic{page}}
\setcounter{figure}{0}
\setcounter{table}{0}
\setcounter{equation}{0}
\setcounter{page}{1}

\begin{center}
\section*{Supplementary Materials for:\\ \vspace{1em} \Large{Scientific Discovery in the Age of \\ AI and Supercomputing}}

\author{
	Stefano Bianchini,
	Aldo Geuna,
	Fazliddin Shermatov
}
\end{center}
\vspace{2em}


\section{Data construction}
\label{sm:data_construction}

In this section, we describe how the datasets used in the study were constructed. We collected data for publications related to high-performance computing (HPC), artificial intelligence (AI), the top 1\% most cited papers, and novelty indicators for the 1\% most cited papers. For the regression analysis, we also collected a set of control variables described below. 

\subsection{HPC data construction}
To identify publications that relied on HPC resources, we analyzed acknowledgment sections and extracted mentions of supercomputing facilities, clusters, and national or institutional HPC centers. This procedure allows us to approximate HPC usage at scale. To assess the validity and completeness of this proxy, we carried out targeted checks using two well-documented HPC infrastructures: the HPC facilities at the University of Strasbourg\footnote{\url{https://hpc.pages.unistra.fr/}} and the HPC4AI platform at the University of Turin\footnote{\url{https://hpc4ai.unito.it/}}. More on the validation in the following.

\subsubsection{Methodology}
Our approach focuses on the acknowledgment section because formal citation practices for research infrastructures, in particular HPC systems, are not yet standardized. Simple keyword searches in titles, abstracts, or main text risk capturing papers that study HPC technologies rather than papers that use them. In contrast, mining acknowledgments allows us to identify publications that relied on HPC resources to conduct their research, regardless of discipline. As shown by prior work \cite{Mayernik2016}, these unstructured statements remain the most reliable source for linking scientific output to the use of supercomputing facilities.

We adopt a deliberately narrow definition of HPC as a technology: a dedicated, high-capacity supercomputing system. Under this definition, general-purpose computing facilities, powerful local hardware, and cloud computing services are excluded. Our keyword set reflects this choice and includes only the following terms: “high performance comput-er/-ing”, “HPC”, “supercomputing”, and “supercomputer”.

\subsubsection{Data retrieval}
Following the methodology defined above, we retrieved data using the Scopus API, querying the acknowledgments field directly via the FUND-ALL() operator. This allowed us to apply our narrowly defined set of case-insensitive keywords and n-grams, requiring an exact match in word order. As of February 2025, this procedure identified 239,298 publications (1963–2025). Among these, approximately 37\% explicitly contain the phrase “high-performance computing”. Table~\ref{tab:hpc_keywords} reports the full frequency distribution of all search terms. A given publication may match multiple keywords.

\begin{table}[htbp]
\centering
\caption{List of HPC keywords and corresponding number of EIDs}
\vspace{1em}
\label{tab:hpc_keywords}
\begin{tabular}{@{}r l r@{}}
\toprule
\textbf{No.} & \textbf{HPC Keyword} & \textbf{Total EIDs} \\
\midrule
1 & high-performance computing & 88,974 \\
2 & supercomputing & 72,035 \\
3 & supercomputer & 53,912 \\
4 & HPC & 48,561 \\
5 & high-performance computer & 12,387 \\
\bottomrule
\end{tabular}
\end{table}

\subsubsection{Validation through two case studies}
\label{sm:data_validation}
To validate the coverage of our keyword-based method, we conducted two detailed case studies using internal user lists from the HPC facilities of the University of Strasbourg (UNISTRA) and the University of Turin (UNITO). We compared the publication output of registered HPC users with the set of papers from the same institutions identified by our acknowledgment-based method. \\

\noindent \textbf{University of Strasbourg}. We obtained the full list of users from the UNISTRA HPC management team (this is the only HPC of the University of Strasbourg). Between 2014 and 2024, a total of 1,405 users received access credentials. We matched these individuals to 843 unique Scopus authors, who collectively produced 10,709 publications in the same period. Using ScienceDirect, we retrieved full texts for 1,859 articles (17.4\% of this output), allowing us to identify which of these papers actually used the HPC (recognizing that not all publications by a user necessarily rely on HPC resources).

To ensure that the full-text subsample was suitable for downstream validation, we first assessed potential sampling bias. Coverage rates were stable across publication years (14–18\% between 2014 and 2024), indicating no systematic time drift. Mean citation counts in the subsample closely matched those of the full corpus, suggesting that the sample does not disproportionately capture high- or low-impact work. Major subject areas showed comparable representation in both sets. Thus, the full-text subset appears sufficiently representative for the purposes of our validation exercise.

We then used the state-of-the-art LLM Gemini 2.0 Flash to inspect the full texts and identify papers that \textit{explicitly} state the use of an HPC resource. Only clear, unambiguous declarations of HPC use were accepted. The model was provided with the following targeted prompt:
\begin{quote}
\begin{sloppypar}
\texttt{Act as a research paper analysis expert focusing on computational resources. Analyze the provided paper text to determine if the authors explicitly state they used a High-Performance Computing (HPC) resource for their research.}

\texttt{Paper text: \{full\_text\}}

\texttt{Instructions:}

\texttt{1. Identify if the paper contains an explicit statement confirming the use of an HPC resource (e.g., HPC facility, supercomputer, computing cluster, specific named HPC center etc.) for the authors' own work described in the paper. Look often in Acknowledgements, Methods, or Computational Details sections.}

\texttt{2. The mention must be unambiguous and directly linked to the execution of simulations, computations, data analysis, or other research tasks presented in the paper. It should state they used the resource.}

\texttt{3. Do not infer usage based solely on the scale of computation described, or general discussion of computational methods without stating where they were run.}

\texttt{Answer the following questions, providing ONLY the answers separated by the | symbol, without any additional text, headers, or explanations:}

\texttt{1. Did the paper explicitly state the use of an HPC resource for the authors' work? (Yes/No)}

\texttt{2. If Yes, provide one direct quote (verbatim sentence or relevant clause containing the explicit statement) from the paper that serves as evidence. If No, write 'No explicit mention found.'.}

\texttt{Provide ONLY the answers separated by the | symbol.}

\texttt{Example Response 1 (HPC Used):}

\texttt{Yes | "The large-scale simulations were performed on the 'Titan' supercomputer located at the Oak Ridge Leadership Computing Facility."}

\texttt{Example Response 2 (HPC Not Explicitly Mentioned or Ambiguous):}

\texttt{No | No explicit mention found.}
\end{sloppypar}
\end{quote}

Under this definition, 195 papers (10.5\%) in the full-text subsample were confirmed by the LLM as having used an HPC resource. Among these 195 HPC-using papers, 137 were also captured by our acknowledgment-based keyword search, yielding a recall rate of 70\%. We manually examined the remaining 58 cases and confirmed that they were genuine HPC-using papers that the keyword search missed because the relevant statements occurred in the methods or results sections, not in the acknowledgments. After applying a minimum-length filter (20 words or 200 characters) to avoid truncated texts, we found no false positives—that is, no papers flagged by the keyword method that the LLM rejected as non-HPC-using. Finally, across novelty, citation impact, and disciplinary composition, we find no statistically significant evidence that the false negatives are concentrated in particular fields or types of papers (Table~\ref{tab:hpc_validation_tests}). \\

\noindent \textbf{University of Turin}. We repeated the same validation procedure for the HPC4AI facility at UNITO. Since its opening in 2020, 111 researchers have been granted access to the facility (the University of Torino has another HPC for which we do not have information). We matched 66 of these users to Scopus author IDs, who together produced 1,477 publications. From this corpus, we retrieved the full texts of 261 papers (17.7\%). As with the UNISTRA sample, we verified that the subset is representative: full-text availability is stable across years, citation patterns closely mirror the full corpus, and the distribution across major subject areas shows no systematic deviation. We therefore consider this sample suitable for further analysis.

The LLM classification identified 21 papers (8\%) that explicitly report using HPC resources. Of these, 13 were also detected by our acknowledgment-based keyword search, corresponding to a capture rate of 62\%. Manual inspection of the remaining 8 papers confirmed that they are all genuine HPC-using articles that escaped keyword detection because the relevant statements appeared outside the acknowledgments. As before, we found no false positives and no evidence of subject-area bias among the false negatives.

These two validation exercises indicate that while the keyword-based approach may not recover every instance of HPC usage, it captures a large majority of cases (62–70\%) with no false positives and without introducing systematic disciplinary distortions. It therefore provides a high-precision, lower-bound estimate of HPC use at scale, appropriate for macro-level analyses of computational practices in science, as in our study.

\begin{table}[ht!]
\centering
\caption{Validation tests: keyword-captured vs.\ missed HPC papers}
\vspace{1em}
\label{tab:hpc_validation_tests}
\begin{threeparttable}
\begin{tabular}{llcc}
\toprule
Dimension & Test & Statistic & $p$-value \\
\midrule
Citation count            & Welch two-sample $t$-test        & $t = 0.752$            & 0.453 \\
Citation count             & Kolmogorov--Smirnov               & $D = 0.152$             & 0.163 \\
Subject area (primary)     & Pearson $\chi^2$                  & $\chi^2 = 20.87$ (df=18) & 0.286 \\
Subject area (all tags)    & Pearson $\chi^2$                  & $\chi^2 = 20.91$ (df=20) & 0.403 \\
Subject area (all tags)    & Permutation (TVD)                 & $\mathrm{TVD} = 0.138$  & 0.394 \\
Novelty (\texttt{new\_word\_d}) & Two-proportion (cont.\ corr.) & $\chi^2 = 0.252$        & 0.616 \\
\bottomrule
\end{tabular}
\begin{tablenotes}[flushleft]
\footnotesize
\item \textit{Notes:} Balance tests comparing keyword-captured HPC papers ($N=150$) against LLM-confirmed HPC papers missed by the keyword method ($N=66$), pooled across the UNISTRA and UNITO validation samples ($N_{\text{gemini}}=216$). None of the six tests reject the null of no difference between the two groups, indicating that missed papers are statistically indistinguishable from captured papers in citation impact, subject-area composition, and novelty, i.e., the keyword method's false negatives do not appear systematically concentrated in specific fields.
\end{tablenotes}
\end{threeparttable}
\end{table}

\subsection{AI data construction}
We construct metadata for AI-related publications using a keyword-based approach. Our initial keyword list combines terms from several established studies on AI measurement \cite{WIPO2019, WIPO2022, Baruffaldi2020, Martinelli2021, Bianchini2022, Bianchini2023}, resulting in 364 candidate terms. To ensure high precision, we further refined this list by removing overly generic expressions and retaining only terms that are specific and central to AI research. The final taxonomy comprises 203 keywords.

Using this curated set, we queried the Scopus API across titles, abstracts, and author keywords. As of March 2025, this procedure identifies 5,193,154 AI-related publications since 2000. The distribution is highly concentrated: the five most frequent keywords account for roughly 50\% of all papers in the dataset (see Table \ref{tab:ai_keywords_two_column}). Note that a publication may contain multiple AI keywords and therefore contribute to several entries in the keyword frequency table.

\begin{center}
\centering
\scriptsize 
\setlength\tabcolsep{4pt} 

\begin{longtable}{@{}r l r | r l r@{}} 

\caption{List of AI Keywords and corresponding number of EIDs} \label{tab:ai_keywords_two_column} \\
\toprule
\textbf{No.} & \textbf{AI Keyword} & \textbf{Total EIDs} & \textbf{No.} & \textbf{AI Keyword} & \textbf{Total EIDs} \\
\midrule
\endfirsthead

\multicolumn{6}{c}{\tablename\ \thetable{} -- Continued from previous page} \\
\toprule
\textbf{No.} & \textbf{AI Keyword} & \textbf{Total EIDs} & \textbf{No.} & \textbf{AI Keyword} & \textbf{Total EIDs} \\
\midrule
\endhead

\midrule
\multicolumn{6}{r@{}}{{Continued on next page}} \\
\endfoot

\bottomrule
\endlastfoot

1 & neural network & 1,116,394 & 2 & machine learning & 772,250 \\
3 & artificial intelligence & 642,564 & 4 & image processing & 579,306 \\
5 & deep learning & 511,800 & 6 & learning algorithm & 305,681 \\
7 & feature extraction & 291,775 & 8 & genetic algorithm & 281,545 \\
9 & pattern recognition & 262,294 & 10 & convolutional neural network & 261,708 \\
11 & computer vision & 249,048 & 12 & vector machine & 231,373 \\
13 & support vector machine & 228,455 & 14 & image segmentation & 222,689 \\
15 & learning model & 203,023 & 16 & virtual reality & 192,361 \\
17 & information retrieval & 188,963 & 18 & decision tree & 151,445 \\
19 & random forest & 144,315 & 20 & natural language processing & 136,759 \\
21 & swarm optimization & 123,150 & 22 & objective function & 113,739 \\
23 & fuzzy logic & 111,195 & 24 & reinforcement learning & 109,158 \\
25 & unmanned aerial vehicle & 107,212 & 26 & speech recognition & 103,386 \\
27 & multi-agent system & 99,315 & 28 & object recognition & 96,557 \\
29 & object detection & 95,927 & 30 & evolutionary algorithm & 94,628 \\
31 & feature selection & 93,599 & 32 & image classification & 93,269 \\
33 & supervised learning & 83,079 & 34 & long short-term memory & 82,250 \\
35 & face recognition & 73,441 & 36 & expert system & 70,514 \\
37 & backpropagation & 65,249 & 38 & multi-objective optimization & 64,585 \\
39 & k-nearest neighbor & 61,818 & 40 & motion planning & 58,671 \\
41 & augmented reality & 58,036 & 42 & transfer learning & 57,058 \\
43 & classification algorithm & 54,866 & 44 & semantic web & 52,131 \\
45 & k-means clustering & 46,458 & 46 & adversarial network & 45,621 \\
47 & generative adversarial network & 43,552 & 48 & sentiment analysis & 41,300 \\
49 & image retrieval & 40,524 & 50 & hidden markov model & 40,368 \\
51 & dimensionality reduction & 37,913 & 52 & random field & 37,289 \\
53 & hierarchical clustering & 35,887 & 54 & knowledge representation & 35,478 \\
55 & active learning & 35,280 & 56 & convolutional network & 34,789 \\
57 & intelligent agent & 34,667 & 58 & autoencoder & 33,336 \\
59 & predictive analytics & 32,548 & 60 & gradient boosting & 31,505 \\
61 & machine vision & 30,808 & 62 & image recognition & 30,603 \\
63 & text mining & 28,282 & 64 & gaussian process & 27,370 \\
65 & emotion recognition & 27,188 & 66 & xgboost & 26,981 \\
67 & bayesian network & 26,737 & 68 & recommendation system & 25,852 \\
69 & independent component analysis & 25,676 & 70 & training dataset & 25,641 \\
71 & gesture recognition & 25,331 & 72 & speech analysis & 24,544 \\
73 & latent variable & 23,904 & 74 & ensemble learning & 23,337 \\
75 & pose estimation & 23,255 & 76 & trajectory tracking & 23,142 \\
77 & clustering analysis & 23,129 & 78 & machine translation & 23,016 \\
79 & swarm intelligence & 22,781 & 80 & image matching & 22,267 \\
81 & gaussian mixture model & 21,760 & 82 & sensor data fusion & 21,087 \\
83 & pattern analysis & 20,497 & 84 & predictive modeling & 20,377 \\
85 & bayesian model & 19,687 & 86 & collaborative filtering & 19,665 \\
87 & obstacle avoidance & 19,660 & 88 & text classification & 18,937 \\
89 & markov decision process & 18,460 & 90 & sparse representation & 18,303 \\
91 & regression tree & 18,001 & 92 & fuzzy c-means & 17,889 \\
93 & autonomous vehicle & 17,681 & 94 & extreme learning machine & 16,863 \\
95 & probabilistic approach & 16,689 & 96 & genetic programming & 16,609 \\
97 & activity recognition & 16,563 & 98 & adaboost & 16,555 \\
99 & feature learning & 15,844 & 100 & q-learning & 15,568 \\
101 & text analysis & 15,185 & 102 & trajectory planning & 15,078 \\
103 & multi-layer perceptron & 14,966 & 104 & self-learning & 14,644 \\
105 & action recognition & 14,295 & 106 & brain-computer interface & 13,865 \\
107 & graphical model & 13,766 & 108 & adaptive learning & 13,284 \\
109 & topic modeling & 13,270 & 110 & face detection & 13,165 \\
111 & computational intelligence & 12,841 & 112 & cognitive model & 12,098 \\
113 & bayesian learning & 12,040 & 114 & multi-task learning & 11,820 \\
115 & mixed reality & 11,807 & 116 & differential evolution algorithm & 11,156 \\
117 & feature engineering & 10,986 & 118 & statistical learning & 10,945 \\
119 & simultaneous localization and mapping & 10,825 & 120 & self-organizing map & 10,253 \\
121 & latent dirichlet allocation & 8,916 & 122 & link prediction & 8,266 \\
123 & distributed processing & 8,224 & 124 & stochastic gradient descent & 8,077 \\
125 & validation dataset & 8,053 & 126 & non-negative matrix factorization & 7,822 \\
127 & dictionary learning & 7,662 & 128 & neuromorphic computing & 7,402 \\
129 & spectral clustering & 7,326 & 130 & text-to-speech & 7,271 \\
131 & nearest neighbor algorithm & 7,177 & 132 & firefly algorithm & 6,783 \\
133 & classification tree & 6,560 & 134 & naive bayes classifier & 6,527 \\
135 & deep belief network & 6,135 & 136 & behavioral analysis & 5,763 \\
137 & target function & 5,723 & 138 & pedestrian detection & 5,640 \\
139 & text recognition & 5,220 & 140 & voice recognition & 5,144 \\
141 & transformer network & 4,944 & 142 & natural language understanding & 4,800 \\
143 & latent semantic analysis & 4,460 & 144 & variational inference & 4,436 \\
145 & boosting algorithm & 4,413 & 146 & memetic algorithm & 3,995 \\
147 & machine intelligence & 3,959 & 148 & autonomous learning & 3,906 \\
149 & handwriting recognition & 3,706 & 150 & kernel learning & 3,672 \\
151 & semantic search & 3,610 & 152 & natural language generation & 3,473 \\
153 & automated reasoning & 3,214 & 154 & stochastic algorithm & 3,170 \\
155 & intelligent machine & 3,161 & 156 & evolutionary computing & 3,104 \\
157 & probabilistic graphical model & 2,986 & 158 & virtual assistant & 2,688 \\
159 & multinomial naive bayes & 2,644 & 160 & inductive logic programming & 2,622 \\
161 & inductive reasoning & 2,587 & 162 & image alignment & 2,585 \\
163 & autonomic computing & 2,552 & 164 & hebbian learning & 2,524 \\
165 & kernel machine & 2,483 & 166 & learning automata & 2,301 \\
167 & convnet & 2,026 & 168 & connectionist model & 1,944 \\
169 & cognitive computing & 1,876 & 170 & relational learning & 1,834 \\
171 & meta learning & 1,795 & 172 & speech-to-text & 1,673 \\
173 & similarity learning & 1,180 & 174 & face analysis & 1,081 \\
175 & blind signal separation & 1,073 & 176 & temporal difference learning & 965 \\
177 & symbolic reasoning & 962 & 178 & factorization machine & 920 \\
179 & policy gradient method & 902 & 180 & instance-based learning & 859 \\
181 & deep forest & 851 & 182 & handwriting analysis & 690 \\
183 & intelligent software agent & 671 & 184 & policy gradient methods & 666 \\
185 & stacked generalization & 643 & 186 & intelligent infrastructure & 524 \\
187 & statistical relational learning & 471 & 188 & conceptual clustering & 470 \\
189 & bootstrap aggregating & 446 & 190 & rule-based learning & 334 \\
191 & intelligent classifier & 330 & 192 & massive parallel processing & 246 \\
193 & artificial reality & 202 & 194 & single linkage clustering & 166 \\
195 & descriptive modeling & 163 & 196 & adaptive boost & 158 \\
197 & cognitive automation & 157 & 198 & trust region policy optimization & 154 \\
199 & intelligence augmentation & 138 & 200 & rankboost & 126 \\
201 & neural turing machine & 76 & 202 & image grammar & 20 \\
203 & totalboost & 11 &  &  &  \\ 

\end{longtable}
\end{center}

\subsection{Top 1\% of publications}
To identify highly cited research, we construct a dataset of the top 1\% most cited papers worldwide. The selection is performed annually and normalized by subject area: a publication is included if its citation count places it within the top 1\% of all papers published in the same year and in the same subject category. Using Scopus data collected on January 18, 2025, we identify 347,153 unique publications published between 2017 and 2024 across 27 subject areas. Within this high-impact corpus, 19\% are classified as AI-related and 1.2\% as HPC-related.

\subsection{Novelty}
We matched the top-1\% corpus with the novelty dataset from \cite{Arts2025sup} using DOIs. The match rate is determined by the temporal coverage of both datasets: the Arts et al. data for 2024 is based on a snapshot from January 9, 2024, while our top-1\% dataset (collected in January 2025) includes all publications from the full 2024 calendar year. After matching, the final sample consists of 281,617 publications.

For each publication, we construct several novelty indicators. Our primary measure is a binary variable equal to 1 if a paper introduces at least one new word. For robustness, we also consider indicators for the presence of new phrases, new word combinations, and new phrase combinations. We also incorporate a semantic-distance-based measure of conceptual novelty, dichotomized into “high novelty” (top quartile of the distribution) versus “lower novelty” (remaining 75\%). These measures allow us to capture both surface-level linguistic novelty and deeper conceptual departures from existing knowledge.

\subsection{Regression control variables}
The novelty sample forms the basis for our regression analyses. Alongside the main variables of interest, we include several controls. Prior citation counts for each paper’s authors were retrieved from the SciVal API and measure cumulative citations accrued before the publication year of the sampled paper. Journal quality is captured using the SCImago Journal Rank (SJR), obtained via the Scopus API. Institutional prestige is measured through the 2024 Academic Ranking of World Universities (Shanghai Ranking), collected from the official website\footnote{\url{https://www.shanghairanking.com/rankings/arwu/2024}}. Institutions are classified into two groups: those belonging to the global top 100 and all others. Summary statistics for all variables used in the regression analysis are provided in Table \ref{tab:summary_statistics}.

\begin{table}[ht] \centering 
  \caption{Summary statistics for regression sample} 
  \label{tab:summary_statistics} \footnotesize\setlength{\tabcolsep}{0.8pt}
  \begin{adjustbox}{max width=\linewidth}
\begin{threeparttable}
\begin{tabular}{@{\extracolsep{5pt}}lcccccccc} 
\\[-1.8ex]\toprule \\[-1.8ex] 
Statistic & \multicolumn{1}{c}{N} & \multicolumn{1}{c}{Mean} & \multicolumn{1}{c}{St. Dev.} & \multicolumn{1}{c}{Min} & \multicolumn{1}{c}{Pctl(25)} & \multicolumn{1}{c}{Median} & \multicolumn{1}{c}{Pctl(75)} & \multicolumn{1}{c}{Max} \\ 
\hline \\[-1.8ex] 
New Word & 281,617 & 0.048 & 0.21 & 0 & 0 & 0 & 0 & 1 \\ 
HPC & 281,617 & 0.011 & 0.11 & 0 & 0 & 0 & 0 & 1 \\  
AI & 281,617 & 0.16 & 0.37 & 0 & 0 & 0 & 0 & 1 \\  
Author Count & 281,617 & 9.8 & 60 & 0 & 3 & 5 & 8 & 7,904 \\
International Collaboration & 280,198 & 0.46 & 0.5 & 0 & 0 & 0 & 1 & 1 \\ 
Prior Citations & 281,538 & 10,234 & 18,299 & 0 & 1,222 & 4,058 & 11,290 & 400,949 \\ 
SCImago Journal Rank 2024 & 191,192 & 3.8 & 5 & 0.1 & 1.3 & 2.2 & 4.1 & 106 \\  
Top-100 Shanghai Rank Dummy & 267,622 & 0.31 & 0.46 & 0 & 0 & 0 & 1 & 1 \\ 
Affiliation Type: University & 281,617 & 0.95 & 0.22 & 0 & 1 & 1 & 1 & 1 \\ 
Affiliation Type: Government & 281,617 & 0.19 & 0.39 & 0 & 0 & 0 & 0 & 1 \\  
Affiliation Type: Industry & 281,617 & 0.13 & 0.34 & 0 & 0 & 0 & 0 & 1 \\
Affiliation Type: Hospital & 281,617 & 0.19 & 0.39 & 0 & 0 & 0 & 0 & 1 \\ 
Affiliation Type: Other & 281,617 & 0.096 & 0.29 & 0 & 0 & 0 & 0 & 1\\  
\bottomrule \\[-1.8ex] 
\end{tabular}
\end{threeparttable}
\end{adjustbox}
\end{table}

\section{Data classification}
\label{sm:data_classification}
\subsection{LLM-based classification of AI-related papers}
We complement our keyword-based AI dataset with a full-text classification approach. This additional layer of analysis allows us to capture cases where AI plays a substantive role in the research even when the relevant keywords do not appear in the title, abstract, or author keyword fields. We focus specifically on the full-text corpus of HPC publications to better characterize how AI methods are used in HPC-intensive research.

Full-text data were obtained through two sources: the ScienceDirect API and the OpenAlex API. Combined, these sources provided full texts for 51\% of our HPC corpus. The distribution of this subsample closely mirrors the overall population across multiple dimensions --subject area, major affiliation countries, author count, citation count, and aggregation type -- with two systematic differences: full-text papers tend to be more recent and more likely to be open-access. Given the absence of meaningful differences along other scientific characteristics, we treat this subsample as sufficiently representative for downstream analyses.

While keyword matching is a well-established strategy in the AI measurement literature, it necessarily offers a surface-level view of a paper’s content. Full-text classification using LLMs provides a more granular understanding of actual AI usage, enabling us to cross-validate keyword-based tagging. Combining the two approaches therefore improves robustness: keywords ensure broad coverage and consistency with existing work, whereas LLM-based reading aids in identifying nuanced or implicit uses of AI that may not be captured at the metadata level.

For the LLM-based classification, we use Gemma2:27B as our main classifier to detect AI-related content in the full-text corpus, with Gemma3:27B used for independent validation. To validate this choice, we used Gemma3:27B to independently review a random sample of 500 papers, comparing outputs on both the AI classification and the subsequent development-vs-use sub-classification. The results converged well: agreement was 93.6\% (468/500) on AI classification and 27/31 on the use-vs-development sub-classification, of which development agreement was 2/3 and use agreement was 25/28. We therefore retained Gemma2:27B as our main classifier.

Across the sample, Gemma2:27B classifies 11\% of papers as AI-enhanced. To evaluate the reliability of these positive classifications, we manually reviewed a random subset of 200 papers flagged as AI-related; of these, 195 (97\%) contained explicit evidence of AI use, providing reassuring evidence of high precision in the LLM-based approach. We then compare the LLM results with the keyword-based method. In the same full-text sample, the keyword approach identifies 9\% of papers as AI-related, a proportion close to what we observe in the full HPC dataset. Across both methods, 8,692 papers are labeled as AI-related.

To quantify agreement, we construct a confusion matrix using the LLM classification as reference. The two methods agree in 93.8\% of cases. Relative to the LLM-based classification (treated as reference), the keyword approach shows a precision of 79.4\%, a recall of 62.1\%, a specificity of 97.9\%, and an F1-score of 69.7\%. This pattern suggests that the keyword method is conservative: it rarely mislabels non-AI papers as AI (2,254 false positives), but it misses a substantial share of AI-related work (5,312 false negatives, more than twice as many as false positives). Cohen’s kappa statistic further confirms substantial agreement between the two methods ($\kappa = 0.7$) (see Table \ref{tab:confusion_matrix}). 

\begin{table}[htbp]
\centering
\caption{Confusion matrix comparing keyword-based and LLM-based classifications}
\vspace{1em}
\label{tab:confusion_matrix}
\begin{tabular}{lcc}
\toprule
 & \textbf{LLM: AI} & \textbf{LLM: non-AI} \\
\midrule
\textbf{Keyword: AI} & 8,692 (True Positive) & 2,254 (False Positive) \\
\textbf{Keyword: non-AI} & 5,312 (False Negative) & 105,275 (True Negative) \\
\bottomrule
\end{tabular}
\end{table}

These discrepancies reflect fundamental differences between the two approaches. The keyword method relies on explicit AI terminology in titles, abstracts, or author keywords, which means it overlooks papers where AI is mentioned indirectly or described in field-specific language. The LLM method, by contrast, analyzes full text and can identify AI-related content even without traditional AI keywords, capturing papers where AI applications are embedded within broader methodological discussions or described using field-specific jargon. The manually validated sample confirms that the LLM approach more accurately identifies AI content, particularly in interdisciplinary work where AI applications may not be apparent from abstracts alone. While keyword methods provide a reliable and high-precision baseline, full-text LLM analysis shows superior performance in detecting AI-related research.

Using the full-text LLM classification, we further divide AI-related papers into two categories: AI-development and AI-use. The first includes research that creates, improves, or studies AI methods, models, algorithms, or theoretical foundations. The second includes research that applies existing AI techniques to address problems in scientific domains outside of AI itself. For the LLM-based classifications, we rely on the following operational definitions:

\begin{itemize}
    \item \textbf{Development}: Research aimed at creating, improving, or studying AI methods, algorithms, models, or theoretical foundations. This can be interpreted as \emph{research on AI}.
    
    \item \textbf{Use}: Research that applies existing AI tools, techniques, or models to solve problems in other domains or applications. This can be understood as \emph{research with AI}.
\end{itemize}

To implement this classification, the LLM was provided with the following prompt:
\begin{quote}
\begin{sloppypar}
\texttt{Act as a research paper analysis expert. Analyze the provided paper text and answer the following questions. Provide ONLY the answers separated by the | symbol, without any additional text.}

\texttt{Paper text: \{full\_text\}}

\texttt{Questions to answer (separate answers with | symbols):}

\texttt{1. Is this paper related to Artificial Intelligence (AI) in any way, either through development, application, or discussion of AI concepts?}

\texttt{   - Instructions: Consider AI as an umbrella term that includes machine learning, deep learning, computer vision, natural language processing, and other subfields, even if the term 'AI' itself is not explicitly mentioned. Classify a paper as AI if it discusses methods, techniques, or applications that fall under these domains.}

\texttt{   - Make no assumptions; provide a deterministic answer. Classify a paper as AI only if there is clear mention or justification in the text. Do not make inferences based on potentially related fields. Focus on explicit and direct mentions of AI.}

\texttt{2. If the answer to Question 1 is 'AI', is the paper about the development or use of AI? (development/use/not\_applicable)}

\texttt{   - Instructions:}

\texttt{     - 'development': Research focused on creating, improving, or studying AI methods, algorithms, models, or theory itself. (Research on AI).}

\texttt{     - 'use': Research that applies existing AI tools, techniques, or models to investigate problems in other domains or applications. (Research with AI).}

\texttt{     - 'not\_applicable': If the paper is not related to AI.}

\texttt{3. Provide a brief justification (maximum one sentence) for your answer to Question 1, citing text from the paper.}

\texttt{Provide ONLY the answers to the questions above, separated by | symbols, without any additional text.}

\texttt{Example response:}

\texttt{AI | development | The paper discusses the development of a new neural network architecture.}

\texttt{non\_AI | not\_applicable | The paper uses pure simulations without explicit AI-related techniques.}
\end{sloppypar}
\end{quote}

Representative examples for each LLM-based classification are provided in Tables \ref{tab:hpc_examples} and \ref{tab:use_development}. For all classification tasks, we rely on deterministic LLM settings to ensure reproducibility of results. Although the LLM method offers richer full-text insights, the keyword-based approach remains the only strategy scalable to the entire corpus. We therefore use the keyword method as our primary classification in the main analysis, and employ the LLM-based results to validate and refine our understanding of AI-related content within HPC publications (e.g., trends in HPC use for core AI vs. applied AI).

\begin{footnotesize} 
\setlength{\tabcolsep}{3pt} 
\begin{longtable}{@{} c | l | c | >{\RaggedRight}p{0.20\linewidth} | >{\RaggedRight}p{0.20\linewidth} | >{\RaggedRight}p{0.20\linewidth} | >{\RaggedRight\hspace{0pt}\arraybackslash}p{0.13\linewidth} @{}}
\caption{HPC applications (with or without AI).} \label{tab:hpc_examples}\\
\toprule
\# & SUB & Use AI & Description & Example of HPC applications & Example of AI applications & DOI \\
\midrule
\endfirsthead
\caption[]{HPC applications (with or without AI) (continued).}\\
\toprule
\# & SUB & Use AI & Description & Example of HPC applications & Example of AI applications & DOI \\
\midrule
\endhead
\midrule
\multicolumn{7}{r@{}}{\footnotesize Continued on next page} \\
\endfoot
\bottomrule
\endlastfoot
1  & PHYS & NO & Modeling the formation and evolution of stellar clusters. & HPC simulates the gravitational interactions of stars within a cluster to understand its dynamics and long-term evolution. & - & \url{10.1051/0004-6361/201833051} \\ \hline
2  & PHYS & YES & Studying the behavior of quantum systems with many interacting particles. & HPC solves complex equations to model the quantum properties of materials and predict new phenomena. & AI enhances molecular dynamics simulations for accurate potential energy modeling & \url{10.1103/PhysRevLett.120.143001} \\ \hline
3  & CHEM & NO & Enhancing gas separation using advanced 2D materials & HPC models molecular dynamics to optimize nanochannel structures & - & \url{10.1038/s41467-017-02529-6} \\ \hline
4  & CHEM & YES & Advancing quantum chemistry with open-source tools & HPC accelerates simulations of electronic structures in complex molecules & AI optimizes quantum chemistry workflows for enhanced efficiency & \url{10.1021/acs.jctc.9b00532} \\ \hline
5  & MATE & NO & Improving solid-state battery performance with garnet electrolytes & HPC models ionic transport to optimize electrolyte interfaces & - & \url{10.1038/nmat4821} \\ \hline
6  & MATE & YES & Accelerating atomistic simulations with machine learning tools & HPC enables large-scale simulations of material properties & AI develops interatomic potentials for faster and accurate modeling & \url{10.1002/adma.201902765} \\ \hline
7  & BIOC & NO & Identifying genetic variants linked to autoimmune diseases & HPC analyzes large-scale genomic datasets to pinpoint disease-associated variants & - & \url{10.1038/s41588-018-0090-3} \\ \hline
8  & BIOC & YES & Using deep learning to classify lung cancer subtypes and predict mutations & HPC processes large-scale histopathology image datasets for training models & AI predicts cancer subtypes and genetic mutations from pathology images & \url{10.1038/s41591-018-0177-5} \\ \hline
9  & ENGI & YES & Enhancing monaural speech quality through advanced spectral mapping & HPC processes large-scale audio datasets for training & AI improves speech intelligibility and quality using gated recurrent networks & \url{10.1109/TASLP.2019.2955276} \\ \hline
10 & ENGI & NO & Tuning electronic properties of 2D materials for advanced technologies & HPC simulates band structure modifications in 2D semiconductors & - & \url{10.1038/s41699-020-00162-4} \\ \hline
11 & CENG & NO & Enhancing lithium-sulfur battery performance with advanced catalysts & HPC models electrochemical reactions to optimize catalyst & - & \url{10.1021/jacs.8b12973} \\ \hline
12 & CENG & YES & Capturing light-field information with achromatic metalenses & HPC simulates light-field data for enhanced imaging systems & AI enhances image reconstruction for high-resolution light-field imaging & \url{10.1038/s41565-018-0347-0} \\ \hline
13 & EART & YES & Leveraging machine learning and big data for precise landslide risk assessment & HPC processes extensive geospatial datasets for high-resolution mapping & AI predicts landslide susceptibility using ensemble machine learning models & \url{10.1016/j.catena.2023.107653} \\ \hline
14 & EART & NO & Enhancing land surface simulations with updated processes and parameterizations & HPC enables large-scale simulations of carbon and nitrogen cycling in ecosystems & - & \url{10.1029/2018MS001583} \\ \hline
15 & COMP & NO & Evaluating protein interactions to understand biological functions & HPC simulates protein docking to predict binding affinities & - & \url{10.1002/wcms.1448} \\ \hline
16 & COMP & YES & Enhancing segmentation accuracy with image-specific fine-tuning & HPC processes large-scale medical imaging datasets for segmentation tasks & AI adapts segmentation models to specific images for improved accuracy & \url{10.1109/TMI.2018.2791721} \\ \hline
17 & ENVI & NO & Exploring vegetation's role in mitigating climate change through biophysical processes & HPC models land-atmosphere interactions to quantify vegetation feedbacks & - & \url{10.1038/nclimate3299} \\ \hline
18 & ENVI & YES & Combining deep learning and process-based models for accurate lake temperature predictions & HPC simulates physical processes in lakes to generate training data & AI predicts depth-specific lake temperatures using hybrid modeling & \url{10.1029/2019WR024922} \\ \hline
19 & ENER & NO & Providing high-resolution solar radiation and meteorological data & HPC processes satellite and surface data for solar energy modeling & - & \url{10.1016/j.rser.2018.03.003} \\ \hline
20 & ENER & YES & Accelerating discovery of battery materials with machine learning techniques & HPC simulates electrochemical properties of battery materials for optimization & AI predicts material properties to design efficient rechargeable batteries & \url{10.1016/j.ensm.2020.06.033} \\ \hline
21 & MATH & YES & Utilizing neural networks to solve partial differential equations efficiently & HPC handles high-dimensional PDE computations for complex systems & AI approximates solutions to PDEs using deep learning techniques & \url{10.1016/j.jcp.2018.08.029} \\ \hline
22 & MATH & NO & Advancing materials modeling with curated pseudopotential libraries & HPC enables high-throughput screening of materials for precision modeling & - & \url{10.1038/s41524-018-0127-2} \\ \hline
23 & MEDI & YES & Revolutionizing pathology with deep learning for large-scale image analysis & HPC processes extensive whole-slide image datasets for model training & AI enables accurate cancer classification and mutation prediction using weakly supervised learning & \url{10.1038/s41591-019-0508-1} \\ \hline
24 & MEDI & NO & Investigating cellular heterogeneity in pulmonary fibrosis using single-cell RNA sequencing & HPC processes large-scale transcriptomic data to identify disease-associated cell populations & - & \url{10.1164/rccm.201712-2410OC} \\ \hline
25 & AGRI & YES & Utilizing time-series Landsat data and machine learning for accurate crop classification & HPC processes extensive Landsat datasets for scalable crop-type classification & AI enhances classification accuracy by integrating spectral and temporal data & \url{10.1016/j.rse.2018.02.045} \\ \hline
26 & AGRI & NO & Investigating how advanced climate and crop models reveal earlier impacts of climate change on agriculture & HPC processes extensive climate and crop simulations for accurate and large-scale agricultural impact predictions & - & \url{10.1038/s43016-021-00400-y} \\ \hline
27 & IMMU & NO & Investigating the diverse roles and states of neutrophils using single-cell RNA sequencing & HPC processes large-scale transcriptomic datasets to map neutrophil heterogeneity & - & \url{10.1038/s41590-020-0736-z} \\ \hline
28 & IMMU & YES & Investigating how macrophage polarization states influence cancer-specific survival in colorectal cancer & HPC processes multiplexed immunofluorescence data for detailed tumor microenvironment & AI identifies macrophage polarization patterns and their prognostic significance & \url{10.1158/2326-6066.CIR-20-0527} \\ \hline
29 & SOCI & NO & Examining the impacts of compound drought and heatwave events on socio-ecosystem productivity and sustainability & HPC simulates climate-hydrology interactions to assess the frequency and severity of compound events & - & \url{10.1038/s41893-022-01024-1} \\ \hline
30 & SOCI & YES & Advancing natural language understanding for Indic languages through curated corpora, benchmarks, and multilingual models & HPC processes extensive linguistic datasets for model training and evaluation & AI develops state-of-the-art multilingual models for Indic languages & \url{10.18653/v1/2023.acl-long.693} \\ \hline
31 & NEUR & NO & Analyzing how affect dynamics contribute to understanding psychological well-being & HPC processes extensive affective time-series data to evaluate emotional patterns and their impact on well-being & - & \url{10.1038/s41562-019-0555-0} \\ \hline
32 & NEUR & YES & Investigating how individual-specific cortical network topography relates to human behavior, cognition, and emotion & HPC processes resting-state fMRI data to map individual-specific cortical networks & AI predicts behavioral phenotypes by analyzing individual-specific network topography & \url{10.1093/cercor/bhy123} \\ \hline
33 & DECI & YES & Creating a new classification and datasets for project scheduling with multi-skilled resources & HPC accelerates the evaluation of scheduling algorithms on large-scale benchmark instances & AI optimizes scheduling algorithms by improving resource allocation and enhancing computational efficiency & \url{10.1016/j.ejor.2022.05.049} \\ \hline
34 & DECI & NO & Providing a comprehensive dataset of climate simulations spanning 540 million years to study Earth's climate evolution & HPC processes extensive climate simulation data to analyze long-term climate patterns & - & \url{10.1038/s41597-022-01490-4} \\ \hline
35 & PHAR & NO & Exploring emerging "next generation" probiotics and their potential therapeutic applications. & HPC processes large-scale microbiome datasets to identify and evaluate next-generation probiotics & - & \url{10.1016/j.jfda.2018.12.011} \\ \hline
36 & PHAR & YES & Developing a deep learning-based method predict short antimicrobial peptides for combating multi-drug resistance & HPC processes genomic sequences to identify potential AMPs efficiently & AI enhances AMP prediction accuracy using convolutional neural networks and optimized feature sets & \url{10.1016/j.omtn.2020.05.006} \\ \hline
37 & BUSI & YES & Evaluating RNNs for time series forecasting & HPC processes large-scale time-series datasets for model evaluation & AI leverages RNNs to improve forecasting accuracy and handle complex temporal patterns & \url{10.1016/j.ijforecast.2020.06.008} \\ \hline
39 & ARTS & YES & Investigating how multitask prompted finetuning enhances crosslingual generalization in multilingual language models & HPC processes multilingual datasets to optimize model finetuning & AI enables multilingual models to generalize across tasks and languages effectively & \url{10.48550/arXiv.2211.01786} \\ \hline
41 & PSYC & NO & Analyzing how affect dynamics contribute to understanding psychological well-being & HPC processes extensive affective time-series data to evaluate emotional patterns and their impact on well-being & - & \url{10.1038/s41562-019-0555-0} \\ \hline
43 & HEAL & YES & Reviewing CNN-based approaches for Alzheimer's disease classification from anatomical MRI & HPC processes large-scale MRI datasets to evaluate CNN architectures for Alzheimer's disease classification & AI enhances classification accuracy by optimizing CNN architectures and addressing reproducibility issues & \url{10.1016/j.media.2020.101694} \\ \hline
45 & ECON & NO & Exploring Arctic sea route navigability due to climate-driven sea ice changes, using high-resolution projections. & HPC processes climate and oceanographic data to model future Arctic navigability & - & \url{10.1016/j.marpol.2015.12.027} \\ \hline
47 & VETE & NO & Evaluating the potential of squid meal and shrimp hydrolysate as sustainable and digestible protein sources for dog nutrition & HPC processes large-scale nutritional datasets to assess the digestibility and metabolic impacts of novel protein sources & - & \url{10.3389/fvets.2024.1360939} \\ \hline
\end{longtable}
\setlength{\tabcolsep}{6pt} 
\end{footnotesize}
\begin{footnotesize}
\setlength{\tabcolsep}{2pt}
\begin{longtable}{@{} c | p{2.5cm} | p{9cm} | p{2.5cm} @{}}
\caption{AI Research: Development vs. Application.} \label{tab:use_development}\\
\toprule
\# & Classification & Paragraph & DOI \\
\midrule
\endfirsthead
\caption[]{AI Research: Development vs. Application (continued).}\\
\toprule
\# & Classification & Paragraph & DOI \\
\midrule
\endhead
\midrule
\multicolumn{4}{r@{}}{\footnotesize Continued on next page} \\
\endfoot
\bottomrule
\endlastfoot
1 & AI\_development & In this work, we take a more radical approach. Instead of hand-crafting a training procedure that emulates aspects of high-level vision pipelines, we embed the feature detector in a complete vision pipeline during training. Particularly, our pipeline addresses the task of relative pose estimation, a central component in camera re-localization, structure-from-motion or SLAM. The pipeline incorporates key point selection, descriptor matching and robust model fitting. We do not need to pre-define ground truth correspondences, dispensing with the need for hard-negative mining. Furthermore, we do not need to speculate whether it is more beneficial to find many matches or few, reliable matches. All these aspects are solely guided by the task loss, i.e. by minimizing the relative pose error between two images. & \url{10.1109/cvpr42600.2020.00500} \\ \hline
2 & AI\_use & To test the hypothesis that the stool microbiome could be used as a reproducible CRC prescreening tool, we performed intra-cohort, cross-cohort and combined-cohort prediction validation on the overall set of 621CRC and control samples using a random forest classifier (Table 1). In intra-cohort cross-validation using species-level taxonomic relative abundances, we observed performances ranging in area under the receiver operating characteristic curve (AUC) score from 0.92 to 0.58, with an average in the deeply sequenced datasets of 0.81 (Fig. 2a). & \url{10.1038/s41591-019-0405-7} \\ \hline
3 & AI\_development & In this paper we present Lazy Resampling, the concepts behind it, and our open-source design and implementation. We show how it improves on existing preprocessing libraries through use of computational geometry at the heart of its design, and the flexibility, simplicity and information preservation that occurs as a result. We show how we innovate from an engineering standpoint by allowing users to take advantage of the power of computational geometry based transforms without them needing to understand any of the mechanisms that underlie its implementation. These contributions serve together to overcome the limitations of existing technologies. We present experiments demonstrating that Lazy Resampling is not only faster, but is able to improve network performance when applied to the training of semantic segmentation networks. & \url{10.1016/j.cmpb.2024.108422} \\ \hline
4 & AI\_development & In this paper, we propose an online hyperparameter tuning method based on population based training (PBT) (Jaderberg et al. 2017). We can then perform hyperparameter adjustment while the AlphaZero algorithm trains, saving precious computing resources. Another significant advantage of using PBT is that this method requires a single run only while incurring a small additional cost for the optimization and evaluation phases of AlphaZero training. & \url{10.1609/aaai.v34i01.5454} \\ \hline
5 & AI\_development & The goal of this paper is to develop an iterative method that predicts four different sets of structural properties: secondary structure, torsion angles, C$\alpha$-atom based angles and dihedral angles and solvent accessible surface area. That is, both local and nonlocal structural information were utilized in iterations. At each iteration, a deep-learning neural network is employed to predict a structural property based on structural properties predicted in the previous iteration. & \url{10.1038/srep11476} \\ \hline
6 & AI\_development & For this task, we developed a machine learning process called "Computer Vision Assisted Feature Enhancer" (CAFE). This approach automatically identifies specific signature features based on their morphology and can specifically discern different types of brain cells and blood micro-vessels. Moreover, CAFE supports parallel computing and can be used on major cloud platforms, such as Hadoop, Amazon Web Service (AWS) and Google Cloud Platform (GCP), to improve scalability. & \url{10.1016/j.physrep.2022.11.003} \\ \hline
7 & AI\_use & Another ET estimates from a different version of the GLEAM model is included in the comparison (GLEAM-HYBRID). This version of the model includes a deep learning algorithm trained with eddy covariance and sap flow measurements to model transpiration stress (Koppa et al., 2022). The temporal and spatial resolutions of this product match the previous version. & \url{10.1016/j.rse.2024.114451} \\ \hline
8 & AI\_use & Using the Human Connectome Project (n = 873, 473 females, after quality control), we directly compared predictive models comprising different sets of MRI modalities (e.g., seven tasks vs. non-task modalities). We applied two approaches to integrate multimodal MRI, stacked vs. flat models, and implemented 16 combinations of machine-learning algorithms. & \url{10.1016/j.neuroimage.2022.119588} \\ \hline
9 & AI\_use & In this study, we analyse the effect of consecutive winters on the arrival delay and the primary delay of the high-speed passenger trains in northern Sweden. Novel statistical learning approaches including inhomogeneous Markov chain model and stratified Cox model were adopted to account for the time-varying risks of train delays and model performance assessment. & \url{10.1016/j.jrtpm.2023.100388} \\ \hline
10 & AI\_use & The SFE of the selected alloys was computed using DFT methods and ML models were then generated, which could also be validated using independent DFT estimates based on SF-defected supercell with suitable configurational averages. These models were then combined with a recently developed model for the intrinsic strength in compositionally complex FCC alloys [36] to explore a much larger alloy space, uncovering a small subset of alloys that are predicted to overcome the strength-ductility trade-off through the exploitation of additional plasticity mechanisms that result from low SFEs. These alloys are located on the Pareto front of the strength-(target SFE) multi-objective space. & \url{10.1016/j.actamat.2021.117472} \\ \hline
\end{longtable}
\setlength{\tabcolsep}{6pt}
\end{footnotesize}

\subsection{Subject areas}
Our primary source for subject-area information is the Scopus All Science Journal Classification (ASJC) system, which assigns publications to 27 top-level scientific fields. For records lacking ASJC codes (most commonly conference proceedings) we complemented this with an LLM-based classification using the Gemma2-27B model, which assigns a field based on the full name of the journal or conference. A random sample of these LLM-generated labels was manually validated and confirmed to be accurate. Consistent with Scopus conventions, publications may belong to more than one subject area.

\subsection{Affiliations}
Author affiliations were classified into five institutional categories: universities, government, industry, hospitals, and other. We implemented a two–step LLM-based procedure using Gemma2-27B and Gemini 2.0. Gemma2-27B was used first to parse and categorize clear, standardized affiliation strings. Gemini 2.0 was then applied to resolve ambiguous or incomplete entries by performing targeted web searches, ensuring high accuracy even for complex institutional names. Affiliations that could not be assigned with confidence were classified under “other”.

To assess the reliability of this approach, we manually reviewed a random sample of 300 institutional names. All 300 were correctly classified. The procedure handled difficult edge cases without error. For example, \textit{Eawag – Swiss Federal Institute of Aquatic Science and Technology}, a government-funded and government-operated research organization, was correctly labeled as government. Likewise, \textit{Ourofino Animal Health}, a private commercial firm in the veterinary pharmaceutical sector, was accurately assigned to industry. The method also reliably handled language variations and institutional complexity.

\section{Statistical models}
\label{sm:methodology}

We model how HPC and AI interact in producing novel, high-impact scientific research. Our goal is to quantify whether each technology independently is correlated to novelty and, more importantly, whether the interaction -joint use- is significant and relevant -papers with a joint use of the two technologies have an higher probability of having a novel content.

To test this, we estimate a probit regression model in which the dependent variable, $\textit{New\ Word}$, equals 1 if a paper introduces at least one new word that is subsequently reused (our primary novelty measure). The main independent variables are two binary indicators capturing the use of HPC and AI, together with their interaction term \textit{$AI \times HPC$}. As shown in Equation \ref{eq:probit_model}, this specification separately identifies the individual correlation of AI and HPC and the additional interaction associated with using both technologies in the same paper.

\begin{equation}
\label{eq:probit_model}
\begin{split}
\mathbb{E}[\text{New\ Word}_{jst} = 1] = & \Phi(\beta_0 + \beta_1 \text{AI}_{jst} + \beta_2 \text{HPC}_{jst}   \\
& + \beta_3 (\text{AI}_{jst} \times \text{HPC}_{jst}) \\
& + \mathbf{X}_{jst}'\gamma + \mathbf{Z}_{jst}'\delta + \varepsilon_{jst})
\end{split}
\end{equation}

\vspace{0.5em}
\noindent where $\mathbf{X}$ includes control variables at the paper, author, and journal level, and $\mathbf{Z}$ includes various fixed effects. We control for the number of authors, international collaboration, prior citations of authors, journal prestige (SCImago Journal Rank), and institutional affiliation type. We also include year fixed effects, subject area fixed effects (27 Scopus ASJC categories), and country fixed effects. Standard errors are clustered at the journal level. 

The regression results are reported in Table \ref{tab:reg_table}. The robustness of these findings is assessed along five dimensions. First, Table \ref{tab:robust_table} reports results using four alternative measures of novelty: new phrase, new word combination, new phrase combination, and semantic distance. Across all specifications, the interaction term between AI and HPC remains positive, and statistically significant in most cases, confirming that papers combining both technologies are systematically more likely to introduce some novelty. Second, Table \ref{tab:robust_keywords} reports results under three alternative AI classification schemes, all benchmarked against the full 203-keyword dictionary. Column (1) restricts to the six most specific of the ten most frequent AI-related keywords (neural network, machine learning, artificial intelligence, deep learning, learning algorithm, computer vision), excluding four broader terms and capturing 45.3\% of AI papers. Column (2) uses the 10 most frequent AI-related keywords (neural network, machine learning, artificial intelligence, image processing, deep learning, learning algorithm, feature extraction, genetic algorithm, pattern recognition, computer vision), capturing 62.1\% of AI papers. Column (3) uses a strict ``core AI'' dictionary of twenty keywords targeting machine learning and AI-specific methodology (artificial intelligence, machine learning, deep learning, neural network, reinforcement learning, supervised learning, transfer learning, active learning, ensemble learning, feature learning, meta learning, multi-task learning, statistical learning, bayesian learning, computational intelligence, machine intelligence, intelligent agent, automated reasoning, symbolic reasoning, cognitive computing), capturing 44.2\% of AI papers. Third, Table \ref{tab:robust_subjects} reports results excluding individual subject areas in turn — Computer Science, Engineering, and Physics — to confirm the relationship is not driven by any single field. Fourth, Table \ref{tab:robust_sample} reports results under additional sample restrictions, excluding top-cited papers, papers with more than 20 authors, and literature reviews. Fifth, Table \ref{tab:reg_table_region} reports results separately by region (United States, European Union, China, and the rest of the world) to assess whether the relationship holds across different geographic and institutional contexts.

\begin{table} 
\centering
\begin{threeparttable}
\linespread{1.0}\selectfont
\centering 
  \caption{Regression results} 
  \label{tab:reg_table} 
\begin{tabular}{@{\extracolsep{3pt}}lccc} 
\\[-1.8ex]\toprule \\[-1.8ex] 
 & \multicolumn{3}{c}{\textit{Dependent variable:}} \\ 
\cline{2-4} 
\\[-1.8ex] & \multicolumn{3}{c}{New Word} \\ 
\\[-1.8ex] & (1) & (2) & (3)\\ 
\hline \\[-1.8ex] 
 AI & 0.228$^{***}$ & 0.186$^{***}$ & 0.181$^{***}$ \\ 
  & (0.020) & (0.024) & (0.023) \\ 

 HPC & 0.131$^{***}$ & 0.114$^{***}$ & 0.120$^{***}$ \\ 
  & (0.037) & (0.042) & (0.043) \\ 
 I(AI $\times$ HPC) & 0.166$^{**}$ & 0.299$^{***}$ & 0.277$^{***}$ \\ 
  & (0.083) & (0.104) & (0.105) \\ 
 SCImago Journal Rank 2024 &  & 0.0002 & 0.001 \\ 
  &  & (0.002) & (0.003) \\ 
 Shanghai Ranking 2024 &  &  & 0.014 \\ 
  &  &  & (0.014) \\ 

 Author Count (log) & 0.221$^{***}$ & 0.222$^{***}$ & 0.218$^{***}$ \\ 
  & (0.011) & (0.015) & (0.015) \\ 
 International Collab. & $-$0.019 & $-$0.006 & $-$0.010 \\ 
  & (0.012) & (0.015) & (0.015) \\ 
 Prior Citations (log) & 0.022$^{***}$ & 0.024$^{***}$ & 0.026$^{***}$ \\ 
  & (0.005) & (0.006) & (0.006) \\ 
 Affiliation type: University & 0.017 & $-$0.009 & $-$0.037 \\ 
  & (0.023) & (0.030) & (0.172) \\ 
 Affiliation type: Government & $-$0.038$^{***}$ & $-$0.030$^{**}$ & $-$0.034$^{**}$ \\ 
  & (0.012) & (0.014) & (0.015) \\ 
 Affiliation type: Industry & 0.097$^{***}$ & 0.077$^{***}$ & 0.072$^{***}$ \\ 
  & (0.017) & (0.019) & (0.019) \\ 
 Affiliation type: Hospital & 0.049$^{***}$ & 0.044$^{*}$ & 0.052$^{**}$ \\ 
  & (0.019) & (0.022) & (0.023) \\ 
 Affiliation type: Other & $-$0.037$^{**}$ & $-$0.029 & $-$0.037$^{*}$ \\ 
  & (0.017) & (0.021) & (0.022) \\ 
 Constant & $-$2.103$^{***}$ & $-$2.105$^{***}$ & $-$2.090$^{***}$ \\ 
  & (0.045) & (0.057) & (0.183) \\ 
\hline \\[-1.8ex] 
Subject Area FE & YES & YES & YES \\
Country FE & YES & YES & YES \\
Year FE & YES & YES & YES \\
\hline \\[-1.8ex] 
Observations & 280,195 & 190,500 & 182,306 \\ 
Log Likelihood & $-$50,379.320 & $-$32,285.870 & $-$30,893.070 \\ 
AIC & 101,052.600 & 64,867.740 & 62,084.150 \\ 
\bottomrule \\[-1.8ex] 
\end{tabular}
\begin{tablenotes}[flushleft]
\footnotesize
\item \textit{Notes:} Cluster-robust standard errors in parentheses (clustered at the journal level). Model (3) is used to predict \textit{New Word} (see main text). Significance level: $^{*}$p$<$0.1; $^{**}$p$<$0.05; $^{***}$p$<$0.01.
\end{tablenotes}

\end{threeparttable}
\end{table}
\begin{table} 
\centering
\begin{threeparttable}
\linespread{1.0}\selectfont
\centering 
  \caption{Alternative regression results} 
  \label{tab:robust_table} 
  \small
\begin{tabular}{@{\extracolsep{3pt}}lcccc} 
\\[-1.8ex]\toprule \\[-1.8ex] 
 & \multicolumn{4}{c}{\textit{Dependent variable:}} \\ 
\cline{2-5} 
\\[-1.8ex] & New Phrase & New Word Comb. & New Phrase Comb. & Semant. Dist. \\ 
\\[-1.8ex] & (1) & (2) & (3) & (4)\\ 
\hline \\[-1.8ex] 
 AI & 0.15$^{***}$ & 0.15$^{***}$ & 0.15$^{***}$ & 0.68$^{***}$ \\ 
  & (0.01) & (0.01) & (0.01) & (0.01) \\ 
 HPC & 0.10$^{***}$ & 0.18$^{***}$ & 0.04 & $-$0.06 \\ 
  & (0.03) & (0.03) & (0.04) & (0.04) \\ 
 I(AI $\times$  HPC) & 0.17$^{**}$ & 0.15$^{*}$ & 0.18$^{*}$ & 0.17$^{**}$ \\ 
  & (0.09) & (0.09) & (0.10) & (0.08) \\ 
 Author Count (log) & 0.19$^{***}$ & 0.21$^{***}$ & 0.14$^{***}$ & $-$0.18$^{***}$ \\ 
  & (0.01) & (0.01) & (0.01) & (0.01) \\ 
 International Collab. & $-$0.02$^{**}$ & $-$0.01 & 0.00 & $-$0.02$^{***}$ \\ 
  & (0.01) & (0.01) & (0.01) & (0.01) \\ 
 Prior Citations (log) & 0.03$^{***}$ & 0.04$^{***}$ & 0.05$^{***}$ & $-$0.02$^{***}$ \\ 
  & (0.00) & (0.00) & (0.00) & (0.00) \\ 
 SCImago Journal Rank 2024 & 0.00$^{***}$ & $-$0.01$^{***}$ & $-$0.01$^{***}$ & $-$0.01$^{***}$ \\ 
  & (0.00) & (0.00) & (0.00) & (0.00) \\ 
 Shanghai Ranking 2024 & 0.03$^{***}$ & $-$0.00 & $-$0.02$^{*}$ & 0.14$^{***}$ \\ 
  & (0.01) & (0.01) & (0.01) & (0.01) \\ 
 Affiliation type: University & 0.00 & 0.01 & $-$0.12 & 0.12 \\
  & (0.11) & (0.10) & (0.10) & (0.13) \\
 Affiliation type: Government & $-$0.02$^{**}$ & $-$0.02$^{*}$ & $-$0.01 & 0.02$^{*}$ \\ 
  & (0.01) & (0.01) & (0.01) & (0.01) \\ 
 Affiliation type: Industry & 0.04$^{***}$ & 0.06$^{***}$ & 0.04$^{***}$ & 0.06$^{***}$ \\ 
  & (0.01) & (0.01) & (0.01) & (0.01) \\ 
 Affiliation type: Hospital & $-$0.02 & $-$0.04$^{***}$ & $-$0.08$^{***}$ & $-$0.16$^{***}$ \\ 
  & (0.01) & (0.01) & (0.01) & (0.01) \\ 
 Affiliation type: Other & $-$0.08$^{***}$ & $-$0.09$^{***}$ & $-$0.09$^{***}$ & 0.10$^{***}$ \\ 
  & (0.01) & (0.01) & (0.01) & (0.01) \\ 
 Constant & $-$1.12$^{***}$ & 0.05 & 0.57$^{***}$ & $-$0.34$^{***}$ \\ 
  & (0.12) & (0.10) & (0.11) & (0.13) \\ 
\hline \\[-1.8ex] 
Subject Area FE & YES & YES & YES & YES \\
Country FE & YES & YES & YES & YES \\
Year FE & YES & YES & YES & YES \\
\hline \\[-1.8ex] 
Observations & 182,306 & 182,306 & 182,306 & 182,306 \\ 
Log Likelihood & $-$84,486.05 & $-$111,811.56 & $-$93,212.66 & $-$87,818.42 \\ 
AIC & 169,270.10 & 223,921.12 & 186,723.32 & 175,934.84 \\ 
\bottomrule \\[-1.8ex] 
\end{tabular}
\begin{tablenotes}[flushleft]
\footnotesize
\item \textit{Notes:} Cluster-robust standard errors in parentheses (clustered at the journal level). Significance level: $^{*}$p$<$0.1; $^{**}$p$<$0.05; $^{***}$p$<$0.01.
\end{tablenotes}

\end{threeparttable}
\end{table}
\begin{table}
\centering
\begin{threeparttable}
\linespread{1.0}\selectfont
\centering
  \caption{Robustness to AI keyword definition}
  \label{tab:robust_keywords}
\small
\begin{tabular}{@{\extracolsep{3pt}}lccc}
\\[-1.8ex]\toprule \\[-1.8ex]
 & \multicolumn{3}{c}{\textit{Dependent variable: New Word}} \\
\cline{2-4}
\\[-1.8ex] & AI Keywords (Top 6) & AI Keywords (Top 10) & AI Keywords (Core AI) \\
\\[-1.8ex] & (1) & (2) & (3)\\
\hline \\[-1.8ex]
 AI & 0.169$^{***}$ & 0.153$^{***}$ & 0.173$^{***}$ \\
  & (0.025) & (0.024) & (0.025) \\
 HPC & 0.129$^{***}$ & 0.124$^{***}$ & 0.132$^{***}$ \\
  & (0.041) & (0.041) & (0.043) \\
 I(AI $\times$ HPC) & 0.228$^{**}$ & 0.244$^{**}$ & 0.210$^{*}$ \\
  & (0.115) & (0.115) & (0.123) \\
 Author Count (log) & 0.237$^{***}$ & 0.237$^{***}$ & 0.237$^{***}$ \\
  & (0.016) & (0.016) & (0.016) \\
 International Collab. & $-$0.011 & $-$0.011 & $-$0.011 \\
  & (0.015) & (0.015) & (0.015) \\
 Prior Citations (log) & 0.023$^{***}$ & 0.023$^{***}$ & 0.023$^{***}$ \\
  & (0.006) & (0.006) & (0.006) \\
 SCImago Journal Rank 2024 & 0.001 & 0.001 & 0.001 \\
  & (0.002) & (0.002) & (0.002) \\
 Shanghai Ranking 2024 & 0.011 & 0.011 & 0.011 \\
  & (0.014) & (0.014) & (0.014) \\
 Affiliation type: University & 0.018 & 0.016 & 0.018 \\
  & (0.158) & (0.158) & (0.158) \\
 Affiliation type: Government & $-$0.038$^{**}$ & $-$0.038$^{**}$ & $-$0.038$^{**}$ \\
  & (0.015) & (0.015) & (0.015) \\
 Affiliation type: Industry & 0.068$^{***}$ & 0.069$^{***}$ & 0.068$^{***}$ \\
  & (0.019) & (0.019) & (0.019) \\
 Affiliation type: Hospital & 0.036 & 0.036 & 0.036 \\
  & (0.024) & (0.024) & (0.024) \\
 Affiliation type: Other & $-$0.036$^{*}$ & $-$0.036$^{*}$ & $-$0.036$^{*}$ \\
  & (0.021) & (0.021) & (0.021) \\
\hline \\[-1.8ex]
Subject Area FE & YES & YES & YES \\
Country FE & YES & YES & YES \\
Year FE & YES & YES & YES \\
\hline \\[-1.8ex]
Observations & 182,272 & 182,272 & 182,272 \\
Log Likelihood & $-$30,843.70 & $-$30,847.10 & $-$30,841.80 \\
AIC & 61,985.40 & 61,992.20 & 61,981.60 \\
\bottomrule \\[-1.8ex]
\end{tabular}
\begin{tablenotes}[flushleft]
\footnotesize
\item \textit{Notes:} Cluster-robust standard errors in parentheses (clustered at the journal level). Column (1) uses a restricted set of 6 core AI keywords (excluding general terms present in the top-10 list). Column (2) uses the top-10 most frequent AI-related keywords. Column (3) uses a strict "core AI" dictionary of 20 keywords targeting machine learning and AI-specific methodology, excluding adjacent computational fields. Significance level: $^{*}$p$<$0.1; $^{**}$p$<$0.05; $^{***}$p$<$0.01.
\end{tablenotes}
\end{threeparttable}
\end{table}
\begin{table}
\centering
\begin{threeparttable}
\linespread{1.0}\selectfont
\centering
  \caption{Robustness to subject area exclusion}
  \label{tab:robust_subjects}
\small
\begin{tabular}{@{\extracolsep{3pt}}lccc}
\\[-1.8ex]\toprule \\[-1.8ex]
 & \multicolumn{3}{c}{\textit{Dependent variable: New Word}} \\
\cline{2-4}
\\[-1.8ex] & No Computer Science & No Engineering & No Physics \\
\\[-1.8ex] & (1) & (2) & (3)\\
\hline \\[-1.8ex]
 AI & 0.145$^{***}$ & 0.163$^{***}$ & 0.190$^{***}$ \\
  & (0.026) & (0.031) & (0.025) \\
 HPC & 0.098$^{**}$ & 0.118$^{**}$ & 0.072 \\
  & (0.044) & (0.046) & (0.056) \\
 I(AI $\times$ HPC) & 0.248$^{*}$ & 0.262$^{**}$ & 0.321$^{***}$ \\
  & (0.130) & (0.104) & (0.120) \\
 Author Count (log) & 0.246$^{***}$ & 0.249$^{***}$ & 0.244$^{***}$ \\
  & (0.017) & (0.017) & (0.017) \\
 International Collab. & $-$0.001 & $-$0.002 & $-$0.010 \\
  & (0.016) & (0.017) & (0.016) \\
 Prior Citations (log) & 0.019$^{***}$ & 0.026$^{***}$ & 0.023$^{***}$ \\
  & (0.006) & (0.007) & (0.006) \\
 SCImago Journal Rank 2024 & 0.001 & $-$0.001 & 0.002 \\
  & (0.002) & (0.002) & (0.003) \\
 Shanghai Ranking 2024 & 0.007 & 0.010 & 0.012 \\
  & (0.015) & (0.016) & (0.015) \\
 Affiliation type: University & $-$0.018 & 0.033 & $-$0.005 \\
  & (0.158) & (0.167) & (0.158) \\
 Affiliation type: Government & $-$0.043$^{***}$ & $-$0.036$^{**}$ & $-$0.040$^{**}$ \\
  & (0.016) & (0.018) & (0.016) \\
 Affiliation type: Industry & 0.068$^{***}$ & 0.086$^{***}$ & 0.086$^{***}$ \\
  & (0.021) & (0.022) & (0.021) \\
 Affiliation type: Hospital & 0.046$^{**}$ & 0.016 & 0.035 \\
  & (0.023) & (0.026) & (0.025) \\
 Affiliation type: Other & $-$0.038$^{*}$ & $-$0.060$^{***}$ & $-$0.050$^{**}$ \\
  & (0.021) & (0.022) & (0.022) \\
\hline \\[-1.8ex]
Subject Area FE & YES & YES & YES \\
Country FE & YES & YES & YES \\
Year FE & YES & YES & YES \\
\hline \\[-1.8ex]
Observations & 162,105 & 143,991 & 164,189 \\
Log Likelihood & $-$26,670.70 & $-$23,456.10 & $-$27,386.70 \\
AIC & 53,637.35 & 47,208.29 & 55,069.40 \\
\bottomrule \\[-1.8ex]
\end{tabular}
\begin{tablenotes}[flushleft]
\footnotesize
\item \textit{Notes:} Cluster-robust standard errors in parentheses (clustered at the journal level). Each column excludes papers from a single subject area from the estimation sample: Computer Science (1), Engineering (2), and Physics (3). Significance level: $^{*}$p$<$0.1; $^{**}$p$<$0.05; $^{***}$p$<$0.01.
\end{tablenotes}
\end{threeparttable}
\end{table}
\begin{table}
\centering
\begin{threeparttable}
\linespread{1.0}\selectfont
\centering
\caption{Robustness to sample restrictions}
\label{tab:robust_sample}
\small
\begin{tabular}{@{\extracolsep{3pt}}lccc}
\\[-1.8ex]\toprule \\[-1.8ex]
 & \multicolumn{3}{c}{\textit{Dependent variable: New Word}} \\
\cline{2-4}
\\[-1.8ex] & Excl. top 0.1\% & Excl. $>$20 authors & Excl. reviews \\
\\[-1.8ex] & (1) & (2) & (3)\\
\hline \\[-1.8ex]
 AI & 0.180$^{***}$ & 0.194$^{***}$ & 0.185$^{***}$ \\
  & (0.023) & (0.024) & (0.025) \\
 HPC & 0.122$^{***}$ & 0.130$^{**}$ & 0.111$^{**}$ \\
  & (0.043) & (0.050) & (0.045) \\
 I(AI $\times$ HPC) & 0.290$^{***}$ & 0.281$^{***}$ & 0.251$^{**}$ \\
  & (0.105) & (0.107) & (0.108) \\
 SCImago Journal Rank 2024 & 0.002 & 0.001 & $-$0.001 \\
  & (0.003) & (0.003) & (0.003) \\
 Shanghai Ranking 2024 & 0.014 & 0.014 & 0.014 \\
  & (0.014) & (0.014) & (0.014) \\
 Author Count (log) & 0.221$^{***}$ & 0.266$^{***}$ & 0.217$^{***}$ \\
  & (0.015) & (0.019) & (0.016) \\
 International Collab. & $-$0.008 & $-$0.014 & $-$0.001 \\
  & (0.015) & (0.017) & (0.016) \\
 Prior Citations (log) & 0.024$^{***}$ & 0.024$^{***}$ & 0.024$^{***}$ \\
  & (0.006) & (0.006) & (0.006) \\
 Affiliation type: University & 0.053 & $-$0.073 & $-$0.035 \\
  & (0.176) & (0.170) & (0.186) \\
 Affiliation type: Government & $-$0.033$^{**}$ & $-$0.024 & $-$0.034$^{**}$ \\
  & (0.015) & (0.016) & (0.016) \\
 Affiliation type: Industry & 0.070$^{***}$ & 0.086$^{***}$ & 0.069$^{***}$ \\
  & (0.019) & (0.021) & (0.020) \\
 Affiliation type: Hospital & 0.055$^{**}$ & 0.037 & 0.045$^{*}$ \\
  & (0.023) & (0.025) & (0.025) \\
 Affiliation type: Other & $-$0.036$^{*}$ & $-$0.024 & $-$0.042$^{*}$ \\
  & (0.022) & (0.025) & (0.023) \\
 Constant & $-$2.179$^{***}$ & $-$2.113$^{***}$ & $-$2.019$^{***}$ \\
  & (0.182) & (0.180) & (0.197) \\
\hline \\[-1.8ex]
Subject Area FE & YES & YES & YES \\
Country FE & YES & YES & YES \\
Year FE & YES & YES & YES \\
\hline \\[-1.8ex]
Observations & 180,643 & 172,785 & 154,363 \\
Log Likelihood & $-$30,413.50 & $-$28,225.00 & $-$28,224.00 \\
AIC & 61,125.00 & 56,748.00 & 56,746.00 \\
\bottomrule \\[-1.8ex]
\end{tabular}
\begin{tablenotes}[flushleft]
\footnotesize
\item \textit{Notes:} Clustered SEs (journal level) in parentheses. Columns exclude, respectively: top 0.1\% by citations, papers with $>$20 authors, and literature reviews. $^{*}$p$<$0.1; $^{**}$p$<$0.05; $^{***}$p$<$0.01.
\end{tablenotes}
\end{threeparttable}
\end{table}
\begin{table}
\centering
\begin{threeparttable}
\linespread{1.0}\selectfont
\centering
\caption{Regression results by region}
\label{tab:reg_table_region}
\small
\begin{tabular}{@{\extracolsep{3pt}}lcccc}
\\[-1.8ex]\toprule \\[-1.8ex]
 & \multicolumn{4}{c}{\textit{Dependent variable: New Word}} \\
\cline{2-5}
\\[-1.8ex] & US & EU & CN & RW \\
\\[-1.8ex] & (1) & (2) & (3) & (4)\\
\hline \\[-1.8ex]
 AI & 0.115$^{***}$ & 0.156$^{***}$ & 0.199$^{***}$ & 0.237$^{***}$ \\
  & (0.033) & (0.033) & (0.032) & (0.047) \\
 HPC & 0.101$^{*}$ & 0.153$^{*}$ & 0.093 & $-$0.149 \\
  & (0.059) & (0.063) & (0.068) & (0.323) \\
 I(AI $\times$ HPC) & 0.323$^{**}$ & 0.286$^{*}$ & 0.095 & 0.642 \\
  & (0.163) & (0.156) & (0.179) & (0.449) \\
 SCImago Journal Rank 2024 & $-$0.0002 & 0.001 & 0.006$^{*}$ & 0.007 \\
  & (0.002) & (0.003) & (0.004) & (0.005) \\
 Shanghai Ranking 2024 & $-$0.005 & $-$0.010 & $-$0.013 & 0.005 \\
  & (0.021) & (0.022) & (0.022) & (0.046) \\
 Author Count (log) & 0.215$^{***}$ & 0.199$^{***}$ & 0.226$^{***}$ & 0.183$^{***}$ \\
  & (0.020) & (0.019) & (0.026) & (0.041) \\
 International Collab. & $-$0.004 & 0.016 & $-$0.077$^{***}$ & $-$0.021 \\
  & (0.025) & (0.023) & (0.022) & (0.035) \\
 Prior Citations (log) & 0.030$^{***}$ & 0.022$^{***}$ & 0.024$^{***}$ & 0.022$^{***}$ \\
  & (0.006) & (0.005) & (0.005) & (0.008) \\
 Affiliation type: University & $-$0.053 & 0.258 & $-$0.112 & 2.964$^{***}$ \\
  & (0.287) & (0.297) & (0.264) & (0.034) \\
 Affiliation type: Government & $-$0.082$^{***}$ & $-$0.046$^{*}$ & $-$0.026 & 0.009 \\
  & (0.025) & (0.025) & (0.020) & (0.051) \\
 Affiliation type: Industry & 0.131$^{***}$ & 0.079$^{***}$ & $-$0.027 & 0.014 \\
  & (0.030) & (0.029) & (0.031) & (0.064) \\
 Affiliation type: Hospital & 0.025 & 0.056$^{*}$ & 0.059 & $-$0.008 \\
  & (0.029) & (0.032) & (0.044) & (0.063) \\
 Affiliation type: Other & $-$0.127$^{***}$ & $-$0.058$^{**}$ & $-$0.085$^{*}$ & $-$0.168$^{*}$ \\
  & (0.028) & (0.028) & (0.050) & (0.102) \\
 Constant & 214.244$^{***}$ & 208.872$^{***}$ & 258.256$^{***}$ & 184.018$^{***}$ \\
  & (9.803) & (10.385) & (10.287) & (15.069) \\
\hline \\[-1.8ex]
Subject Area FE & YES & YES & YES & YES \\
Country FE & NO & NO & NO & NO \\
Year FE & YES & YES & YES & YES \\
\hline \\[-1.8ex]
Observations & 58,108 & 70,018 & 66,006 & 29,215 \\
Log Likelihood & $-$10,978.80 & $-$11,780.20 & $-$12,510.60 & $-$4,112.30 \\
AIC & 22,439.60 & 24,050.40 & 25,109.20 & 8,238.60 \\
\bottomrule \\[-1.8ex]
\end{tabular}
\begin{tablenotes}[flushleft]
\footnotesize
\item \textit{Notes:} Cluster-robust standard errors in parentheses (clustered at the source/journal level). Columns correspond to region-specific samples: US, EU, CN (China), and RW (rest of world). Significance level: $^{*}$p$<$0.1; $^{**}$p$<$0.05; $^{***}$p$<$0.01.
\end{tablenotes}
\end{threeparttable}
\end{table}
\section{Inequality in supercomputing capacity}
\label{sm:computing_capacity}

We use the \textit{Top500} list of supercomputers to construct a proxy for national compute capacity. Compute capacity is measured in floating-point operations per second (FLOP/s), which quantify the number of arithmetic operations a processor can execute per second. Following convention, we rely on the reported peak performance of each system (in FLOP/s) to measure annual and cumulative compute capacity by country. For more details, see the official list at \url{https://top500.org/lists/top500/}. Using these data, we construct a dataset that covers 56 countries over more than 30 years. 

Initial descriptive patterns reveal substantial global disparities in computational capacity. As of 2023, the top four countries in the Top500 list account for 74\% of all listed supercomputers and of total installed compute capacity. Although the geographical coverage of the list has broadened over time, with the number of represented countries rising sharply until the mid-2010s and stabilizing at 56 countries today, the distribution of computing power remains heavily skewed.

To illustrate this trend, the top 5\% of countries held 58\% of global computational capacity in 1993; by 2023, their share had increased to 65\%, a rise of 7 percentage points. This indicates that, despite broader participation, the bulk of compute resources continues to accumulate within a small set of technologically advanced countries.

To quantify these disparities, we employ the Generalized Entropy Index with parameter 0 (GE(0)), also known as the Mean Log Deviation (MLD). MLD is particularly sensitive to changes in the lower tail of the distribution, making it well suited to detecting inequality among countries with modest or emerging computational capacity. The Generalized Entropy Index with parameter 0 is: 

\begin{equation}
GE(0) = \frac{1}{N} \sum_{i=1}^{N} \ln\left(\frac{\mu}{y_i}\right)
\end{equation}

\vspace{0.5em}

\noindent where $N$ is the number of countries, $y_i$ is the computational capacity for country $i$, and $\mu$ is the mean computational capacity across countries. MLD ranges from 0 (perfect equality) to higher positive values indicating greater inequality.


\newpage

\section{Other figures}
\label{sm:other_figures}

\begin{figure}[ht!]
    \centering
    \includegraphics[width=\textwidth]{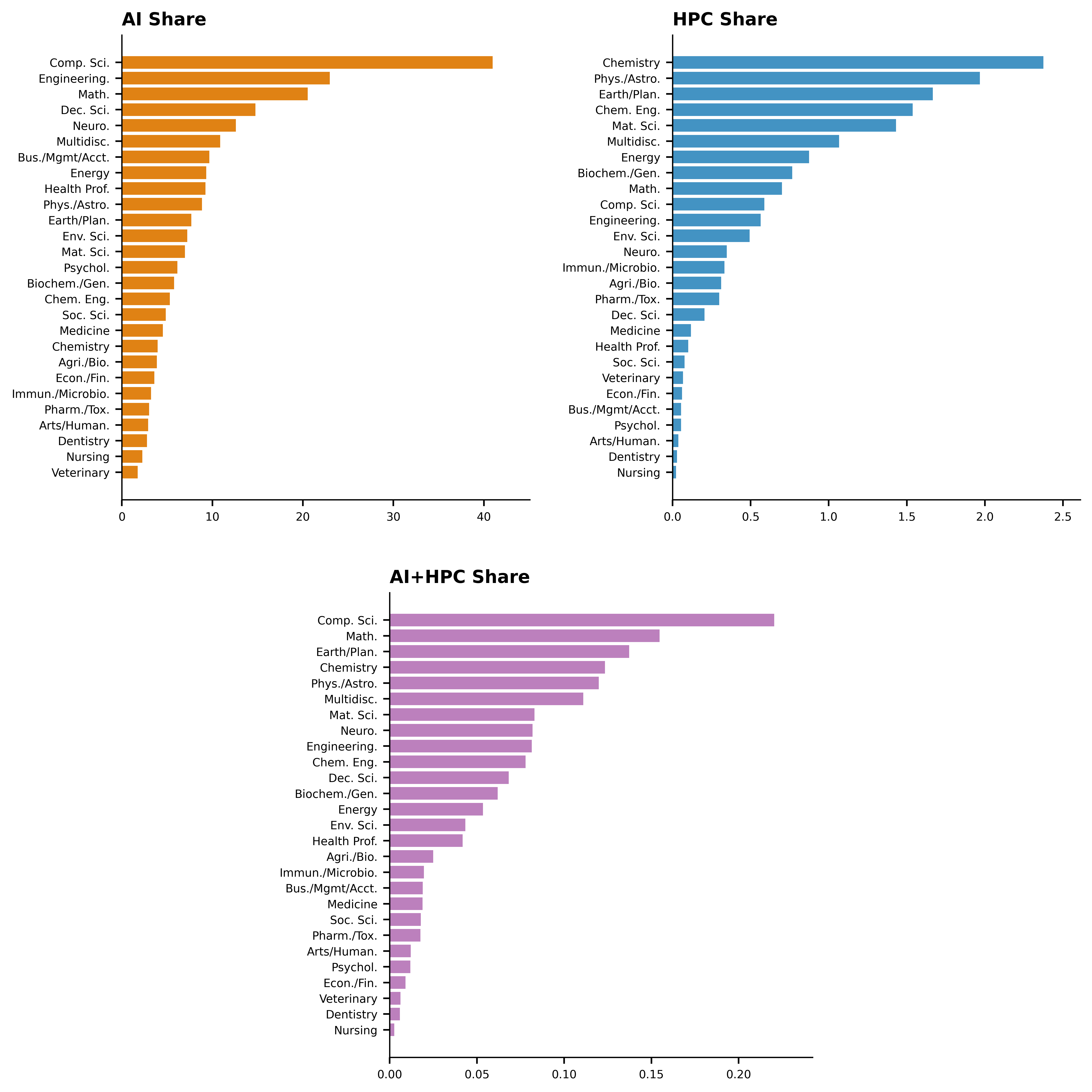} 
    \caption{\textbf{Convergence of AI and HPC in scientific research (all fields).}
    Disciplinary distribution of AI, HPC, and combined AI+HPC work. For each field, bars show the share of total output involving AI.}
    \label{fig:1sm}
\end{figure}

\begin{figure}[ht!]
    \centering
    \includegraphics[width=\textwidth]{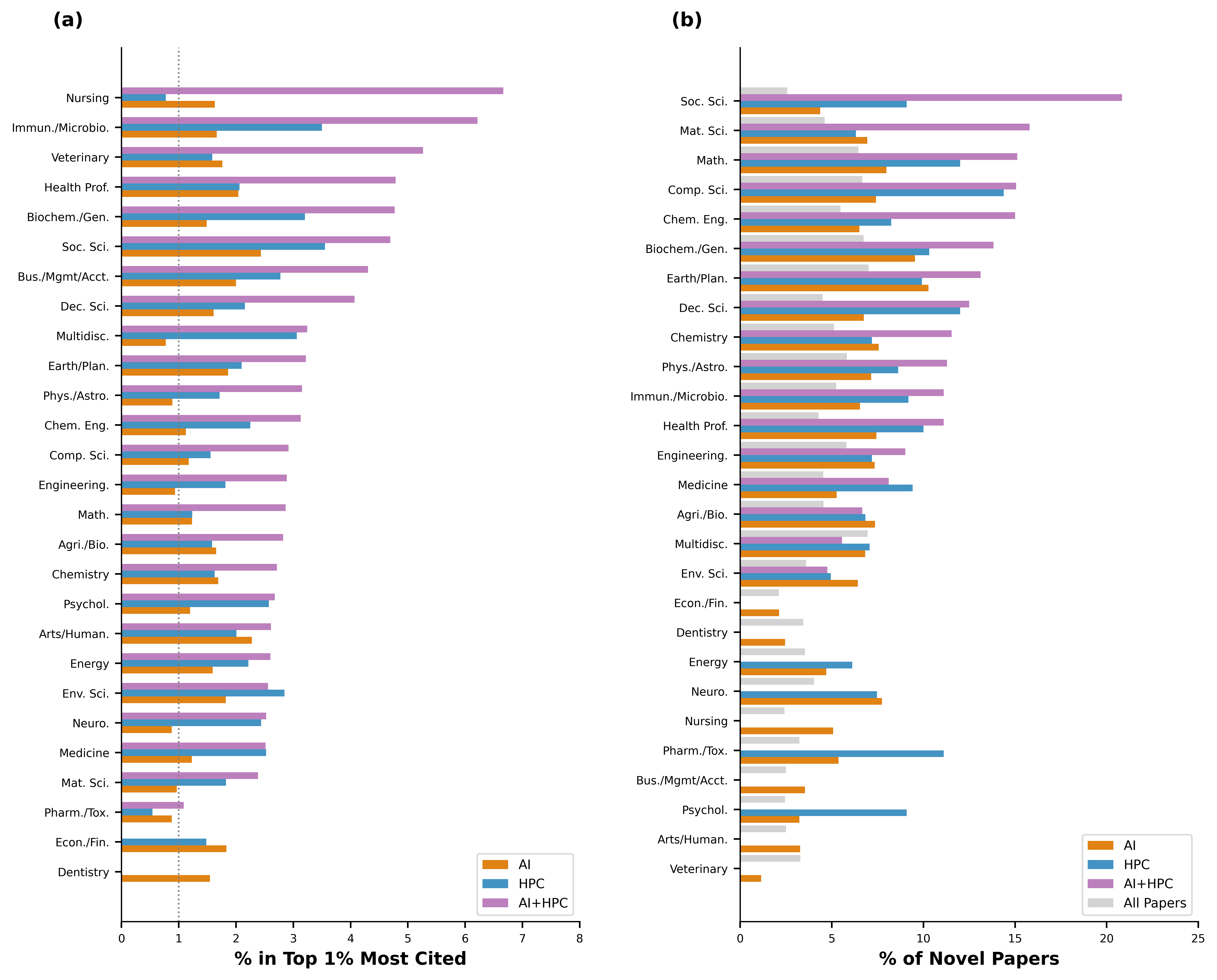}
    \caption{\textbf{Breakthrough potential in AI and HPC-powered science.}
    (\textbf{A}) Share of papers in the top 1\% of citations by field and computational approach. The dashed line marks the 1\% baseline expected by random chance. 
    (\textbf{B}) Share of novel papers (i.e., those introducing a new term subsequently reused) among the top 1\% most cited, across fields and computational approaches.
    In fields where AI+HPC papers remain relatively few, estimates should be interpreted with caution. Results for fields with larger sample sizes are shown in the main text.}
    \label{fig:2sm}
\end{figure}

\begin{figure}[ht!]
    \centering
    \includegraphics[width=\textwidth]{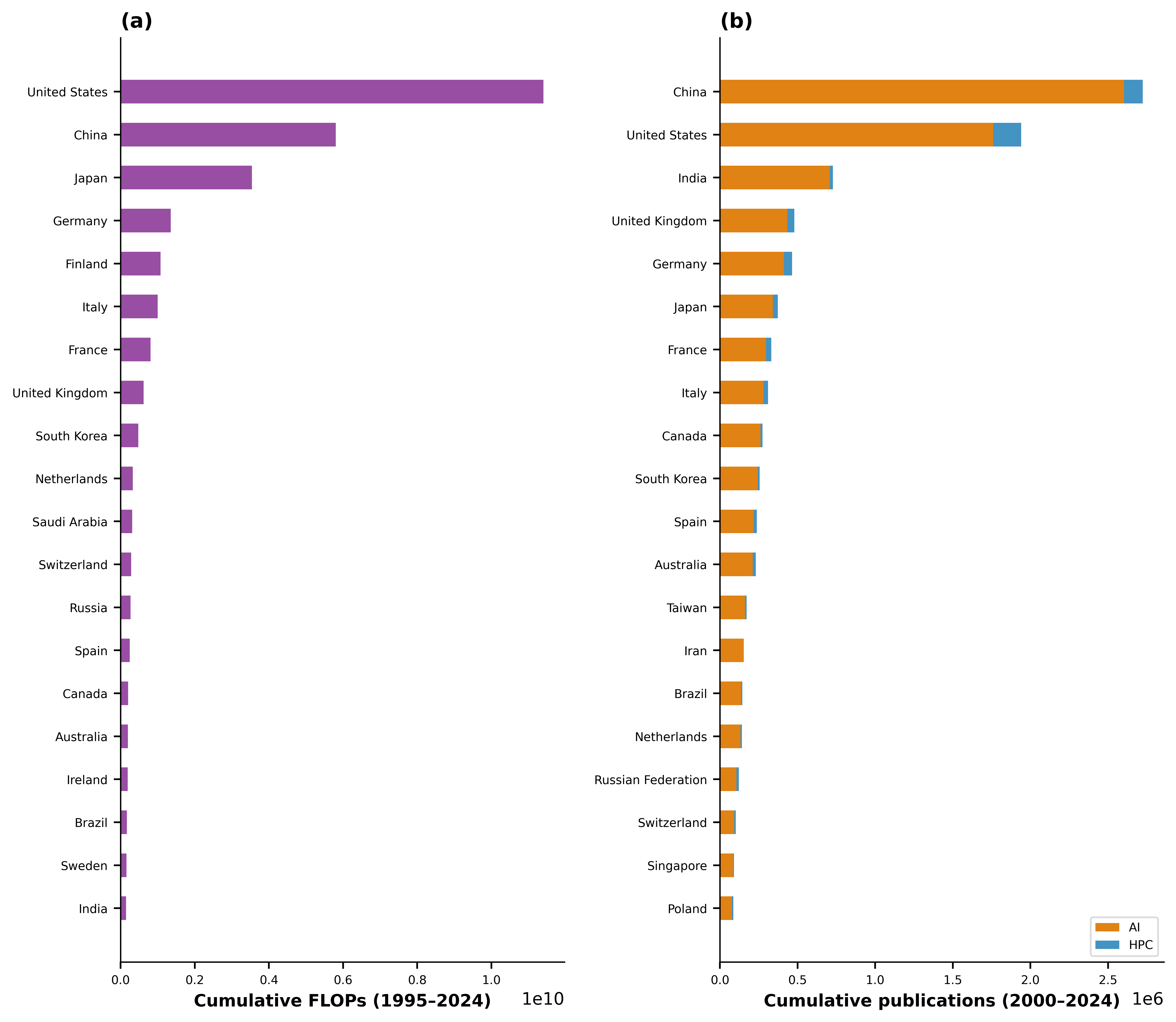}
\caption{\textbf{Global inequality in computational resources and AI/HPC-powered science.}
(\textbf{A}) Top 20 countries by cumulative supercomputing capacity (FLOPs), 1995–2024.
(\textbf{B}) Top 20 countries by cumulative AI and HPC publications, 2000–2024.}
    \label{fig:3sm}
\end{figure}

\begin{figure}[ht!]
    \centering
    \includegraphics[width=\textwidth]{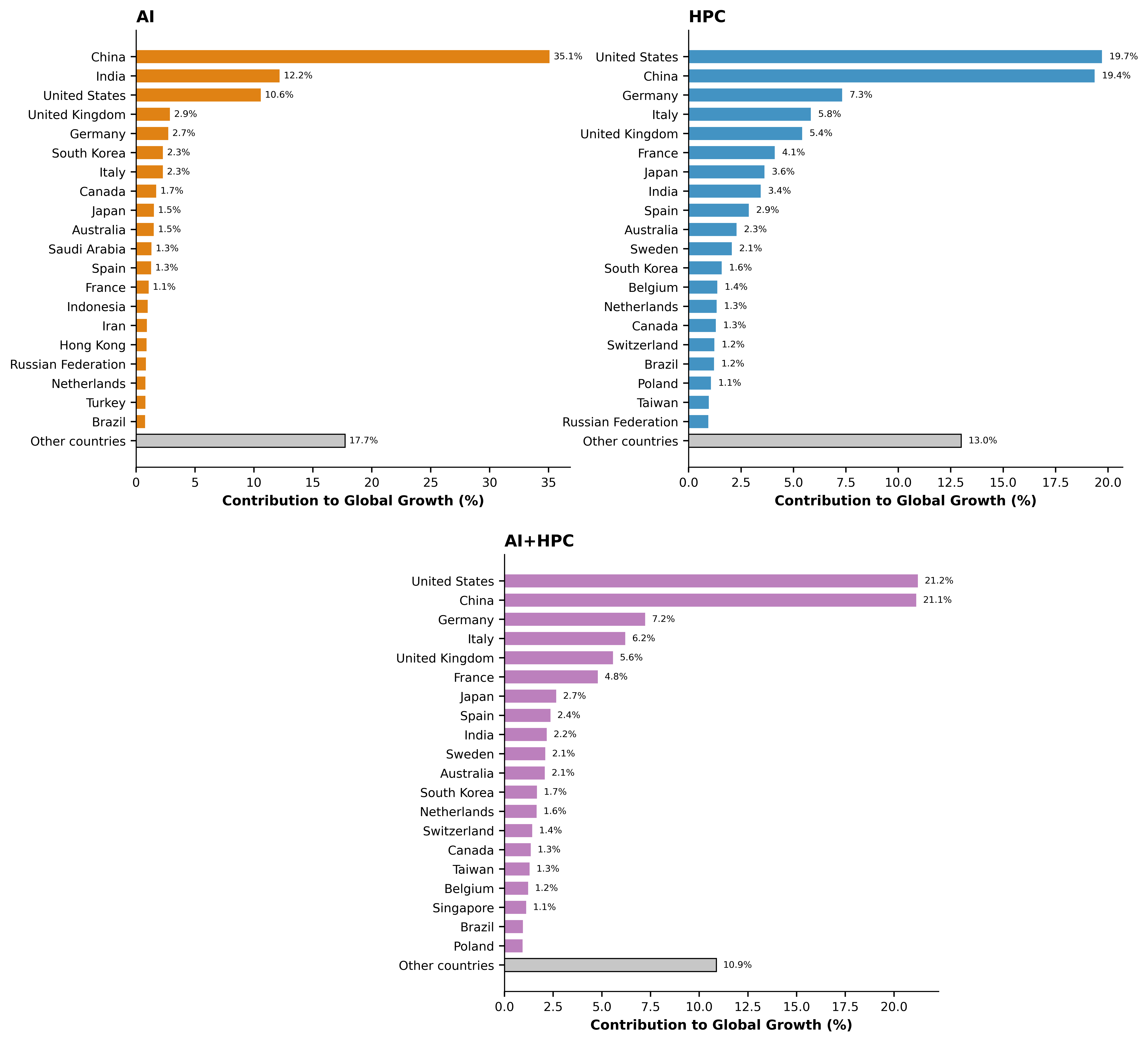}
\caption{\textbf{Contribution of individual countries to the global growth in AI-, HPC-, and AI+HPC-related publications}
Bars report each country's share of the worldwide increase in publications between 2014 and 2024, the period during which cross-country inequality increased most rapidly. “Other countries” aggregates all countries outside the top 20 contributors.}
    \label{sm:figure_sm4}
\end{figure}

\clearpage
\newpage

\renewcommand{\refname}{Supplementary references}

\clearpage

\end{document}